\documentclass[12pt,preprint]{aastex}

\shortauthors{Carciofi, Bjorkman}
\shorttitle{Non-Isothermal Solutions for
Keplerian Disks}

\begin{document}

\title{NLTE Monte Carlo Radiative Transfer:  II. Non-Isothermal Solutions for
Viscous Keplerian Disks 
}

\author{A. C. Carciofi}
\affil{Instituto de Astronomia, Geof\'{i}sica e Ci\^encias Atmosf\'ericas, Universidade de S\~ao Paulo, 
Rua do Mat\~ao 1226, Cidade Universit\'aria, 05508-900, S\~ao Paulo, SP, BRAZIL}
\email{carciofi@usp.br} 
\and
\author{J. E. Bjorkman}
\affil{Ritter Observatory, M.S. 113, Dept. of Physics and Astronomy,
University of Toledo, Toledo, OH 43606-3390} 

\begin{abstract}

We discuss the basic hydrodynamics that determines the density structure of the disks around hot stars.  Observational evidence supports the idea that these disks are Keplerian (rotationally supported) 
gaseous disks.  A popular scenario in the literature, which naturally leads to the formation of Keplerian disks, is the viscous decretion model. 
According to this scenario, the disks are hydrostatically supported in the vertical direction, while the radial structure is governed by the viscous transport. This suggests that the temperature is one primary factor that governs the disk density structure.
In a previous study we demonstrated, using 3-D NLTE Monte Carlo simulations, that viscous keplerian disks can be highly non-isothermal. In this paper we build upon our previous work and solve the full problem of the steady-state non-isothermal viscous diffusion and vertical hydrostatic equilibrium.
We find that the self-consistent solution departs significantly from the analytic isothermal density, with potentially large effects on the emergent spectrum.
This implies that non-isothermal disk models must be used for a detailed modeling of Be star disks.
\end{abstract}


\section{Introduction \label{introduction}}

Disks are observed in a wide variety of astrophysical systems, and they are created in both inflows and outflows whenever the rotation speed greatly exceeds the radial flow speed.  Examples of disk systems formed 
during infall include: young stellar objects, mass transfer binaries, and
active galactic nuclei.  Similarly, disks also form in the outflowing 
winds of luminous stars, such as Be stars, B[e] stars, and possibly LBV and 
AGB stars.  From the kinematic point of view, the disks of hot stars are probably relatively simple, 
so they form ideal laboratories for studying the physics of outflowing circumstellar disks.  For a general review of Be stars and their disks, see \citet{por03}.

The current paradigm for Be star disks is that of a geometrically thin viscous
Keplerian disk in vertical hydrostatic equilibrium (\citealt{lee91}; \citealt{oka01}).
This paradigm has been corroborated by several recent studies \citep[e.g.][]{car06b,mei07}.
%The viscous decretion scenario (\citealt{lee91}; \citealt{oka01}) is the current paradigm to explain the properties of circumstellar disks of Be stars. 
%For instance, recent theoretical results demonstrated that this scenario successfully explain the observations of the Be star $\delta$ Scorpii \citep{car06b}.
The time-independent solution for the viscous decretion disk structure is relatively straightforward to obtain if one assumes the disk to be isothermal (e.g. \citealt{bjo05}; \citealt{oka01}). In this case, one finds that the disk is in vertical hydrostatic equilibrium and because the rotation velocity is much larger than the sound speed, the disk is geometrically very thin, with opening angles of a few degrees only.
In addition, the solution of the viscous outflow results in a simple power-law for the radial dependence of the gas density.

In a previous paper \citep[][hereafter Paper I]{car06a}, we used the three-dimensional (3-D), non-local thermodynamic equilibrium (NLTE) code HDUST to study the temperature structure of Keplerian disks.
We found that those disks can be highly non-isothermal, mainly in their denser inner parts. Later, our results were largely confirmed by \citet{sig07}, who studied the thermal properties of the Be star $\gamma$ Cas.

In Paper I we adopted the isothermal analytic density structure for the disk. 
This is clearly inconsistent with the fact that the disks are non-isothermal, and it is the purpose of this paper to remove this inconsistency. 
We present the first self-consistent solutions for the structure of viscous decretion disks, taking into full account all the relevant physical processes involved. This is a somewhat complicated problem, because the disk temperature controls the geometry (via the hydrostatic equilibrium and viscous diffusion equations), which in turn determine the heating and, hence, the temperature of the disk \citep{ken87}. Solving this problem only became feasible after the advent of large computer clusters.

We also study the emergent spectra of representative model and compare them with those of isothermal models to understand to what extent the non-isothermal solution is required for fitting actual observations.

\section{Disk Theory}

The viscous decretion model is essentially the same as that employed for protostellar disks (e.g., \citealt{bjo97}), the primary difference being that Be disks are outflowing, while pre-main-sequence disks are inflowing. 

It is supposed that some (as yet unidentified) mechanism injects material with Keplerian velocities into the base of the disk. 
Provided this material injection is continual, eddy/turbulent viscosity then transports angular momentum from the inner boundary of the disk outwards. 

The derivation of the density structure for isothermal disks is well-known. Below, we solve this problem for the non-isothermal case.
We follow the derivation of \citet{bjo05}, assuming that the temperature is axisymmetric, i.e., $T=T(\varpi,z)$.

\subsection{Hydrostatic Structure}  

Let us, initially, ignore viscosity. In cylindrical coordinates ($\varpi,\phi,z$), the steady-state fluid equations are
\begin{eqnarray}
   \frac{1}{\varpi}\frac{\partial}{\partial\varpi}(\varpi\rho v_\varpi)
 + \frac{1}{\varpi}\frac{\partial}{\partial\phi}  (\rho v_\phi)
 +                 \frac{\partial}{\partial z}    (\rho v_z)
  &=& 0 \enspace , \label{eq:continuity} \\
   v_\varpi \frac{\partial v_\varpi}{\partial \varpi}  
 + \frac{v_\phi}{\varpi} \frac{\partial v_\varpi}{\partial \phi}
 + v_z \frac{\partial v_\varpi}{\partial z}
 - \frac{v^2_\phi}{\varpi}
  &=& - \frac{1}{\rho} \frac{\partial P}{\partial \varpi} 
 + f_\varpi \enspace , \label{eq:varpi_momentum}\\
   v_\varpi \frac{\partial v_\phi}{\partial \varpi}  
 + \frac{v_\phi}{\varpi} \frac{\partial v_\phi}{\partial \phi}
 + v_z \frac{\partial v_\phi}{\partial z}
 + \frac{v_\varpi v_\phi}{\varpi}
  &=& - \frac{1}{\rho \varpi} \frac{\partial P}{\partial \phi} 
 + f_\phi \enspace , \label{eq:phi_momentum}\\
   v_\varpi \frac{\partial v_z}{\partial \varpi}  
 + \frac{v_\phi}{\varpi} \frac{\partial v_z}{\partial \phi}
 + v_z \frac{\partial v_z}{\partial z}
  &=& - \frac{1}{\rho} \frac{\partial P}{\partial z} 
 + f_z \enspace .\label{eq:z_momentum}
\end{eqnarray}
where $\rho$ is the gas density, $P$ is the pressure and $v_\varpi$, $v_\phi$ and $v_z$ are the velocity components. 
In the absence of any other force, the external force components are given by the gravity of the central star
\begin{eqnarray}
    f_\varpi &=& -\frac{GM\varpi}{(\varpi^2+z^2)^{3/2}} \enspace , \\
    f_z      &=& -\frac{GM z}{(\varpi^2+z^2)^{3/2}}     \enspace .
\end{eqnarray}
Assuming circular orbits ($v_\varpi=0$, and $v_z=0$), the only non-trivial fluid equations are
\begin{eqnarray}
    \frac{1}{\rho} \frac{\partial P}{\partial \varpi} 
          &=& \frac{v^2_\phi}{\varpi} + f_\varpi \enspace , 
                                \label{eq:varpi_hydrostatic} \\
    \frac{1}{\rho} \frac{\partial P}{\partial z} 
          &=& f_z \enspace , 
                                \label{eq:z_hydrostatic}
\end{eqnarray}
the $\varpi$- and $z$-momentum equations, respectively.  
To specify the pressure, we introduce the equation of state $P=a^2\rho$ ,where $a=a(\varpi,z)$ is the sound speed that is related to the gas kinetic temperature by $a=(kT)^{1/2}(\mu m_{\rm H})^{-1/2}$. In this last expression, $k$ is the Boltzmann constant, $\mu$ is the gas molecular weight and $m_{\rm H}$ is the mass of the hydrogen atom.
. %,  which we assume to be axisymmetric.

In the thin-disk limit ($z \ll \varpi$), we obtain
\begin{eqnarray}
    v_\phi &=& V_{\rm crit} \left({R/\varpi}\right)^{1/2} \enspace , \label{eq:vphi}\\
       \frac{\partial \ln (a^2\rho)}{\partial z} 
  &=& -\frac{V^2_{\rm crit}Rz}{a^2 \varpi^3}
       \enspace , \label{eq:hseq}
\end{eqnarray}
where the critical velocity, $V_{\rm crit} \equiv \left(GM/R\right)^{1/2}$, is the Keplerian orbital speed at 
the stellar surface and $R$ is the stellar radius.

The above equations mean that the disk rotates at the Keplerian orbital speed and is hydrostatically supported in the vertical direction.  Defining the disk scaleheight
\begin{equation}
H_0(\varpi) \equiv \frac{a(z=0)}{V_{\rm crit}}\varpi^{3/2}
\enspace ,
\label{eq:scale_height}
\end{equation}
 we obtain, by integrating  the hydrostatic equilibrium equation (\ref{eq:hseq}) along the $z$-direction and writing the result in terms of the temperature
\begin{equation}
\ln \left[\frac{T(\varpi,z)\rho(\varpi,z)}{T_0(\varpi)\rho_0(\varpi)}\right] =
-\frac{1}{2H_0^2(\varpi)} \int_0^{z^2} \frac{T_0(\varpi)}{T(\varpi,z^\prime)}d{z^\prime}^2
\;,
\end{equation}
where the subscript 0 indicates values for $z=0$. Solving for $\rho(\varpi,z)$ we obtain 
\begin{equation}
\rho(\varpi,z) = \rho_0(\varpi)\frac{T_0(\varpi)}{T(\varpi,z)} 
\exp \left[ 
-\frac{1}{2H_0^2(\varpi)} \int_0^{z^2} \frac{T_0(\varpi)}{T(\varpi,z^\prime)}d{z^\prime}^2
 \right]
 \;.
\label{eq:rho01}
\end{equation}

As we shall see below, it is useful to express $\rho(\varpi,z)$ in terms of the disk surface density, $\Sigma(\varpi)$, written as
\begin{equation}
    \Sigma(\varpi)=\int_{-\infty}^\infty \rho(\varpi,z) \, dz \enspace .
\label{eq:sigmadef}
\end{equation}

Substituting eq.~(\ref{eq:rho01}) into eq.~(\ref{eq:sigmadef}) and solving for $\rho_0(\varpi)$ we obtain
\begin{eqnarray}
\rho(\varpi,z) = \frac{\Sigma(\varpi)}{T(\varpi,z)}
\exp \left[ 
-\frac{1}{2H^2_0(\varpi)} \int_0^{z^2} \frac{T_0(\varpi)}{T(\varpi,z^\prime)}d{z^\prime}^2
 \right]
\\
\times
\left\{
\int_{-\infty}^{\infty}
\frac{1}{T(\varpi,z)}
\exp \left[ 
-\frac{1}{2H^2_0(\varpi)} \int_0^{{z}^2} \frac{T_0(\varpi)}{T(\varpi,z^{\prime})}{dz^{\prime}}^2
 \right]
dz
\right\}^{-1}
 \;.
\label{eq:rho}
\end{eqnarray}

\subsection{Viscous Outflow} 

Clearly, in eq.~(\ref{eq:rho}) the disk density scale $\Sigma(\varpi)$ is completely undetermined, because for a pure Keplerian disk we can choose to put an arbitrary amount of material at a given radius, $\varpi$.
To determine the density scale, we must include a mechanism that transports material from the star outwards. The most popular way to generate decretion is to consider viscous effects.

If we add viscosity to the fluid equations (\ref{eq:continuity}) to (\ref{eq:z_momentum}), we can still assume that the disk is axisymmetric and that the vertical structure is 
hydrostatic ($v_z=0$).
However, the presence of an outflow implies that $v_\varpi 
\ne 0$. 
The $\varpi$- and $z$-momentum equations are the same as before, so $v_\phi$ 
and $\rho$ are the same as in the pure Keplerian case [eqs. (\ref{eq:vphi}) and (\ref{eq:rho})].

The continuity equation~(\ref{eq:continuity}) becomes
\begin{equation}
     0=\frac{\partial}{\partial \varpi} (2 \pi \varpi \Sigma v_\varpi)
                                                                \enspace ,
\label{eq:disk_continuity}
\end{equation}
so the decretion rate ${\dot M} = 2 \pi \varpi \Sigma v_\varpi$ is a
constant (independent of $\varpi$).  The %radial velocity component is
viscous outflow speed is 
given by
\begin{equation}
            v_\varpi = \frac{\dot M}{2 \pi \varpi \Sigma} \enspace .
\label{eq:radvel}
\end{equation}
The remaining fluid equation is the $\phi$-momentum equation 
(\ref{eq:phi_momentum}).  This equation now is more complicated because 
viscosity exerts a torque, which is described by the viscous shear stress 
tensor, $\pi_{ij}$.  Including this shear stress, the $\phi$-momentum 
equation becomes
\begin{equation}
  v_\varpi \frac{\partial v_\phi}{\partial \varpi} 
     + \frac{v_\varpi v_\phi}{\varpi}
   = \frac{1}{\rho \varpi^2}
     \frac{\partial}{\partial \varpi} (\varpi^2 \pi_{\varpi\phi}) \enspace 
,
\label{eq:navier_stokes_phi_momentum}
\end{equation}
where
\begin{equation}
    \pi_{\varpi \phi} = \nu \rho \varpi 
          \frac{\partial (v_\phi/\varpi)}{\partial \varpi}
     \enspace .
 \end{equation}
For the \emph{kinematic viscosity}, $\nu$, we use the prescription introduced by \citet{sha73}, according to which $\nu=\alpha a H$, where $0 < \alpha < 1$. The  viscous shear stress 
tensor then becomes
 \begin{equation}   
    \pi_{\varpi \phi} = 
        -\frac{3}{2}\alpha a^2 \rho
           \enspace .
\label{eq:shear_stress}
\end{equation}
Multiplying eq.~(\ref{eq:navier_stokes_phi_momentum}) by $\rho \varpi^2$ and 
integrating over $\phi$ and $z$, we find
\begin{equation}
    {-\dot M} \frac{\partial}{\partial \varpi}(\varpi v_\phi)
    =  \frac{\partial {}}{\partial \varpi} 
  \int_{-\infty}^\infty \varpi \pi_{\varpi \phi} \,
                 2 \pi \varpi \, dz   
  \enspace .
\label{eq:jdot_gradient}
\end{equation}
Substituting eq.~(\ref{eq:shear_stress}) into eq.~(\ref{eq:jdot_gradient}) and integrating it over $\varpi$ we obtain
\begin{equation}
    {\dot M} \varpi v_\phi
    =  3\pi \alpha  \varpi^2 a_0^2
  \int_{-\infty}^\infty \frac{T(\varpi,z)}{T_0(\varpi)} \rho(\varpi,z) dz 
  + \rm constant
  \enspace ,
\label{eq:bla}
\end{equation}
where the integration constant is determined by the outer boundary condition of the disk.
Finally, defining the weighted mass average temperature
\begin{equation}
\langle T(\varpi) \rangle \equiv
\frac{
\int_{-\infty}^{\infty} T(\varpi,z) \rho(\varpi,z) dz
}{
\int_{-\infty}^{\infty} \rho(\varpi,z) dz
}
=
\frac{
\int_{-\infty}^{\infty} T(\varpi,z) \rho(\varpi,z) dz
}{
\Sigma(\varpi)
}
\enspace ,
\label{eq:avT}
\end{equation}
substituting eq.~(\ref{eq:avT}) into eq.~(\ref{eq:bla}), and solving for the surface density we find
\begin{equation}
    \Sigma(\varpi)=\frac{\mu_0 m_{\rm H} {\dot M} V_{\rm crit} R^{1/2}}
           {3 \pi k \alpha \langle T(\varpi) \rangle \varpi^{3/2}}
           \left[(R_{\rm d}/\varpi)^{1/2}-1\right] \enspace ,
\label{eq:disk_Sigma}
\end{equation}
where $\mu_0$ is the mid-plane mean molecular weight, and the \emph{disk outer radius}, $R_{\rm d}$, is determined by the integration constant in eq.~(\ref{eq:bla}).
Physically this arises from the outer boundary condition that sets the sonic location (critical point) in the disk outflow. If the disk is truncated by a physical mechanism, such as tidal truncation by a binary, this location will be near the truncation radius of the disk.

We have now completed the hydrodynamic description of a non-isothermal viscous decretion disk.
To fully determine the problem, one must specify the decretion rate, $\dot{M}$, the disk radius, $R_{\rm d}$, and the disk temperature structure, $T(\varpi, z)$. 
%Note that the mean molecular weight is also a function of position.

\subsection{Properties of the Isothermal Solution}

The analytical solution for the isothermal case is fully discussed elsewhere (e.g. \citealt{bjo05}; \citealt{oka01}), but since we will focus below on the comparison between the isothermal and non-isothermal cases, it is useful to present a brief outline of its properties.

Assuming a constant temperature we find, from eq.~(\ref{eq:disk_Sigma}), that the surface density depends on radius as
\begin{equation}
    \Sigma(\varpi) \propto \varpi^{-3/2} \left[(R_{\rm d}/\varpi)^{1/2}-1\right]
\enspace .
\end{equation}
For large disks ($R\ll R_{\rm d}$), the surface density becomes a simple power-law, $\Sigma(\varpi)\propto \varpi^{-2}$.

The isothermal disk flares quite dramatically with radius, $H\propto\varpi^{1.5}$. 
Since $\rho \propto \Sigma/H$, we obtain that the isothermal disk density profile is quite steep, $\rho \propto \varpi^{-3.5}$.

%Another important property of isothermal Keplerian disks can be readily derived from Eq. (\ref{eq:scale_height}). Since $v_\phi$ is much larger than the sound speed (the former is of the order of several hundreds of km/s and the later is a few tens of km/s), the disk scaleheight is small compared to the stellar radius, i.e. the disk is geometrically thin.

Finally, we find that, for large disks,  the radial velocity is a
linear function of the radial distance, $v_\varpi\propto\varpi$.

%{\bf Here: discuss the effects of $\alpha$ on the disk structure? If so, the text below may be a starting point}
%To conclude this brief description of the properties of isothermal Keplerian viscous disks, let us analyze how varying the $\alpha$ parameter of \citet{sha73} affects the disk structure. In the formalism of viscous decretion disks, viscosity plays the role of transferring momentum from the inner parts of the disk to the outer parts. In other words, without viscosity the fluid elements would orbit the star forever and have $v_\varpi = 0$. The larger the parameter $\alpha$, the more efficient is the momentum transfer and the larger is the outflow velocity. Therefore, for a given $\dot{M}$, a larger $\alpha$ results in smaller surface densities.
%In our modeling, changing $\alpha$ from 0.1 to 1, which is the usual range of this parameter in the literature, affects only the nominal value of $\dot{M}$, because the change in the velocity structure (by a factor of 10) has little effects on the macroscopic properties of the gas (temperature and hydrogen level populations). Thus, we arbitrarily adopt 0.1 as the value for $\alpha$ throughout the paper.

\section{Self-consistent Hydrostatic Solutions}

In this section we determine numerically the solution of non-isothermal viscous decretion disks, and study how it compares with the simple analytical solutions of the isothermal case. 
For this, we modified the computer code HDUST described in Paper I.

HDUST is a fully 3-D NLTE Monte Carlo radiation transfer code that solves the radiative equilibrium temperature for arbitrary density and velocity configurations. In brief, the NLTE Monte Carlo simulation performs a full spectral synthesis by emitting stellar photons with random frequencies sampled using a Kurucz model atmosphere for the central star \citep{kur94}.  Each photon is followed as it travels through the envelope (where it may be scattered, or absorbed and reemitted, many times) until it escapes.  
%When the photons escape, they are binned according to their emergent direction and frequency, which gives the emergent spectrum.

During the simulation, whenever a photon scatters, it changes direction, Doppler shifts, and becomes partially polarized. Similarly, whenever a photon is absorbed, it is not destroyed; it is reemitted locally with a new frequency and direction determined by the local emissivity, $j_\nu$, of the gas.  Note that we include both continuum processes and spectral lines in the opacity and
emissivity of the gas.  Since photons are never destroyed (absorption is always followed by reemission of an equal energy photon packet), our procedure automatically enforces radiative equilibrium and conserves flux exactly \citep[see][]{bjo01}.

The interaction (absorption) of the photons with the gas provides a direct sampling of all the radiative rates, as well as the heating and cooling of the free electrons.  Consequently, we adopt an iterative scheme in which at the end of each iteration we solve the rate equations to update the level populations and electron temperature of the gas. The iteration proceeds until convergence of all state variables is reached.

In paper I we discussed in great detail the solution for the temperature structure of a viscous decretion disk with a fixed density structure.
We found that the disk temperature is a hybrid between that of optically thick disks of Young Stellar Objects and optically thin winds from Hot Stars.  
For the optically thin upper layers, as well as for the outer disk  ($\varpi/R \gtrsim 10$) , the temperature is nearly isothermal, with an average temperature of about 60\% of $T_\mathrm{eff}$. 

At the mid-plane the temperature has a complex structure.  Close to the photosphere the temperature is a very steep function of $\varpi$, with a profile well described by a flat blackbody reprocessing disk \citep{ada87},
\begin{equation}
T_{\rm d}(\varpi)  = 
%\left(
\frac{T_\star}{\pi^{1/4}}
%\right)^{1/4}
\left[
\sin^{-1}\left( \frac{R}{\varpi} \right) - 
%\left(
\frac{R}{\varpi}
%\right) 
\sqrt{1-\frac{R^2}{\varpi^2}}
\right]
^{1/4},
\label{eq:als}
\end{equation}
where $T_\star$ is the temperature of the radiation that illuminates the disk. 
When the disk becomes optically thin vertically, the temperature departs from this curve and rises back to the optically thin radiative equilibrium temperature. 
%, which is approximately constant in the winds of Hot Stars. 

For the present work we modified HDUST to solve simultaneously for both the disk density structure and the gas state variables.
At the end of each iteration, we first determine the new temperature and, then,  solve the hydrostatic equilibrium equation, eq.~(\ref{eq:hseq}), for the vertical structure of the disk, as well as the steady outflow, eq.~(\ref{eq:disk_Sigma}), for  surface density.
The new density thus obtained is used for the next iteration.
The procedure is repeated until the density, temperature, and level populations converge (this typically requires between 5 to 30 iterations, depending on the disk density: the larger the density, the larger the number of iterations).

The primary parameter determining the disk structure is the density scale, which is set by the ratio of the stellar mass loss rate and the parameter  $\alpha$ [Eq.~(\ref{eq:disk_Sigma})]. The other parameters, such as stellar effective temperature, stellar radius, etc., albeit important, do not change significantly the properties of the temperature solution (Paper I).
For this reason, we choose a fixed set of parameters to define the star and the disk size and leave only the density scale as a free-parameter in this study. 
We define 3 density regimes: low mass loss rate, high mass loss rate and an equal mass case, whose meaning will be explained below. 
All fixed model parameters are listed in table~\ref{tab:fixed} and the parameters associated with each density regime are listed in table~\ref{tab:models}.

For the parameter $\alpha$ we arbitrarily choose a value of 0.1 so that
the viscous diffusion timescale roughly agrees with the observed disk
dissipation and creation timescales \citep{oka01}.  However, the actual value
of $\alpha$ is, within a reasonable range, largely irrelevant for the
present problem.  This becomes evident when one analyses the role of $\alpha$ in the disk dynamics.
From Eqs.~(\ref{eq:radvel}) and (\ref{eq:disk_Sigma}) we find that $v_\varpi \propto \alpha$; hence, $\alpha$ controls the viscous outflow speed.
However, as long as the outflow is highly subsonic, which is true for values of $\alpha$ in the range 0.1 --- 1, for instance, the disk dynamics are entirely dominated by rotation.
It follows that if we vary
$\alpha$ within the above range while adjusting $\dot{M}$ to maintain
a constant value of $\dot{M}/\alpha$ (so that the disk density does
not change), all the properties of the solution will be essentially
unaltered.

Our solutions for the disk structure are shown in Figures~\ref{fig:density} to \ref{fig:structure}.
In Figure~\ref{fig:density} we compare the density structure of the consistent numerical solution with the structure of a corresponding isothermal disk.
For the later model, we adopted a value for the gas kinetic temperature of 60\% of $T_\mathrm{eff}$, or $12\,000\;\rm K$ (see Paper I).
Two things are evident from the figure: the consistent solution has smaller scaleheights and larger densities in the mid-plane than the isothermal density. 
Those features are easier to see in Figure~\ref{fig:deltarho}, in which we plot the ratio between the self-consistent and isothermal solutions, $\rho/\rho_\mathrm{iso}$. 
For the high mass loss rate case (bottom plots) the differences in the inner part of the disk are quite significant: the equatorial densities are enhanced by a factor of up to $\approx3$, whereas the upper layers of the disk have much lower densities. 
For low mass loss rates there is a slight enhancement in the equatorial regions, but the differences are much smaller than the previous case.

These results can be understood in terms of the temperature structure. In the last section we demonstrated that what controls the steady flow solution is $\langle T \rangle$, the weighted mass average of the temperature, {eq.~(\ref{eq:avT}). This quantity is shown in Figure~\ref{fig:temperature} for both the low and high mass loss rate models.
For the later, $\langle T \rangle$ is large at the base of the star ($\approx16\,000\;\rm K$), drops very quickly to values close to $6\,000\;\rm K$ around $\varpi/R = 5$ and reaches a near-constant plateau of $\approx12\,000\; \rm K$ around $\varpi/R = 10$. The low values of $\langle T \rangle$ cause the material to collapse towards the equator, thus increasing the mid-plane densities and decreasing the densities of the upper layers.
Note that $\langle T \rangle$ is very similar to the mid-plane temperature (left plots of Figure~\ref{fig:temperature}), because the mid-plane region dominates the integral in eq.~(\ref{eq:avT}).

For low densities there are also important variations in $\langle T \rangle$, but they are confined to much smaller radii and are of smaller amplitude. For this reason, the consistent solution does not differ significantly from the corresponding isothermal model for low mass loss rates, as shown in Figure~\ref{fig:deltarho}.

In Figure~\ref{fig:structure} we exhibit other aspects of the self-consistent solution for the disk density.
We define the disk opening angle, $\theta$, as the angle subtented by one scaleheight at a given radius,
\begin{equation}
\theta(\varpi) = \sin^{-1}\left[\frac{H_0(\varpi)}{\varpi}\right]. \label{eq:opangle}
\end{equation}
Let us first analyze the opening angle of the high mass loss rate model (Fig.~\ref{fig:structure}, top right panel). Initially, at $\varpi/R = 1$, the scaleheight is larger than the isothermal value because of the large temperature at the base of the disk. As the temperature quickly falls, \emph{the disk flares negatively}, i.e., its geometrical thickness becomes smaller with growing radius. This is a subtle effect but quite interesting in view of the striking contrast with the isothermal solution.
The disk begins to flare again around $\varpi/R = 2$ as a result of the decrease of gravitational potential, but it does not flare nearly as much as an isothermal disk. 
When $\langle T \rangle$ reaches its minimum and starts rising again, the disk flares dramatically, reaching the isothermal opening angle when $\langle T \rangle$ becomes approximately constant. 

These remarkable properties of the disk opening angle are also present in the low density model, but are much less pronounced. En passant, we note that the opening angle for the inner disk is very small for all models, in agreement with the general belief that the disks of Be stars are geometrically very thin. 

The remaining plots of Figure~\ref{fig:structure} illustrate different aspects of the radial density structure.
%The properties of the solution for the radial density structure is also illustrated in Figure~\ref{fig:structure}. 
The plots second from bottom, for instance, show in a more quantitative way the density enhancement of the mid-plane, which can be larger than 2.5. In those plots, a feature that is hard to see in Figure~\ref{fig:deltarho} is quite evident: very close to the star there is a \emph{density decrement} in the mid-plane, which arises as a result of the large temperatures there. 
As before, the effects are less conspicuous for low mass loss rates. 

Another striking feature of the solution can be seen in the bottom panels of Figure~\ref{fig:structure}, where we plot $d \ln(\rho_0)/d \ln(\varpi)$ as a function of $\varpi$. This quantity gives the local index of the density profile, were it to be fitted by a power-law. The radial dependence of the density departs \emph{very significantly}  from a simple power-law. Close to the star ($\varpi/R \lesssim 3$) the density profile is much flatter than the isothermal profile, but after the $\langle T \rangle$ turnover point, which for the high mass loss rate model is around $\varpi/R = 3$,  the profile becomes significantly steeper. The density profile only approaches the isothermal profile after $\langle T \rangle$ reaches the near constant plateau at $\varpi/R \approx 10$.

We end this section discussing a subtle property of the solution for the disk structure. For a given $\dot{M}$, the total disk mass of the consistent solution is always larger than the disk mass of an isothermal disk with $T_e = 60\%\;T_\mathrm{eff}$.
For instance, for the high mass loss rate case ($\dot{M} = 5\times10^{-11}\; M_\sun \rm\;yr^{-1}$) the disk mass is $3.6\times10^{-9} \;M_\sun$ whereas the isothermal disk mass is $2.8\times10^{-9} \;M_\sun$.
The total disk mass scales with the global weighted mass average of the temperature, which is lower than the assumed temperature for the isothermal model.
%This is due to the fact that the the inner parts of the disk are significantly cooler than the assumed temperature of the isothermal model. 
For this reason, we defined the equal mass cases in Table~\ref{tab:models} so that the total disk mass of the consistent numerical model equals that of the isothermal model. 
%The equal mass case will be used in the next section, where we investigate the observational signatures of the consistent disk solution.
 
\section{Observational Signatures}

In the previous section we have shown that the self-consistent solution for the disk structure can depart significantly from the analytical isothermal solution.
Here we study what are the effects of the density changes on the emergent spectrum. 
For each density regime we compare models with the same total disk mass.

To clearly separate between different physical effects, i.e., what is the effect of the density and what is the effect of the temperature, we define three different models. The first model, which we call \emph{consistent model}, corresponds to the full non-isothermal solution for both the hydrostatic equilibrium and radial outflow. The second model, which we call \emph{mixed model}, has a fixed density structure given by the analytical isothermal solution, but the gas state variables (temperature and level populations) are determined by HDUST. Those are the same type of models presented in Paper I. Finally, we define \emph{isothermal models}, for which both the density and temperature are fixed, and only the level populations are calculated.

The comparison between the mixed and isothermal models allows us to pinpoint the effects of the temperature structure on the emergent spectrum; on the other hand, the comparison between the consistent and mixed models enables us to determine the effects of the density changes.

Based on the previous description of the properties of the solution, we expect the largest differences between the three models to occur for higher mass loss rates (higher densities). 
Indeed, Figures~\ref{fig:mosaic1} and \ref{fig:mosaic2}  confirm this expectation, since the results for the low mass loss rate models are nearly identical, whereas the results for higher densities can be substantially different, mainly for the emission line strengths and the continuum polarization.

It is important to emphasize that, for the low mass loss rate regime, the results for the isothermal model only match those of the mixed and consistent models because the proper choice of the temperature was made: $12\,000\;\rm K$ or 60\% of $T_\mathrm{eff}$. If another temperature choice had been made, the differences would be significant. 
We can conclude, therefore, that models with low mass loss rates and with the correct choice of the temperature do a very good job in determining the emergent spectrum of viscous decretion disks. This means that employing the more complex mixed or consistent solutions is not necessary (of course, the NLTE treatment of the level populations must still be done).

The situation for the high mass loss rate case is the opposite, since relevant differences exist in the emergent spectrum of the three models. The key for understanding those differences resides, in part, in the Hydrogen level populations. 
As shown in Figure~\ref{fig:lev}, at the non-isothermal part of the disk ($\varpi/R \lesssim 10$) the level populations can be different by factors of up to 2 --- 3.
This indicates that both the temperature structure and the density structure affects the populations in a complex way.
% the consistent and mixed models have, in general, similar values for the populations, throughout the disk. but the values in the mid-plane can be different by a factor of 2. 
%The same is not true for the isothermal models, however, for which the excitation structure is very different than the structure of the mixed and consistent models.
 
Let us first compare the emergent spectra of the mixed and isothermal models, which, as stated above, can give us clues on how the temperature structure affects the results.
We recall that those models have exactly the same density distribution. %, and the only difference lies in the electron temperature. 
There are significant differences in the continuum polarization (factor of about 2) and in the H$\alpha$ emission (about 50\%), and those differences come directly from the dissimilar excitation structure in the non-isothermal part of the disk.
This proves the importance of using the correct solution for the temperature and a full NLTE treatment of the populations.
In addition, % to the differences in the polarization and H$\alpha$, 
there are small differences in the IR spectrum: the slope is slightly flatter for the isothermal model, which results in larger fluxes for long wavelengths. This is a direct result of the mid-plane temperature structure, a subtle effect which was fully discussed in Paper I.

The differences between the mixed and consistent models are also significant; the consistent model produces smaller polarization levels and H$\alpha$ emission than the mixed model. 
 %, but in this case the results are not so easily understood. 
Those models differ mainly in the density structure, but there are also slight differences in temperature and level populations as a result of the density changes.
%The consistent model produces smaller polarization than the mixed model. 
%This is, at least in part, a result of the geometrical changes in the inner disk, which is much thinner for the consistent model. The decrease in the geometrical thickness translates into less stellar flux being scattered by the disk and, hence, to smaller polarization levels.

%The consistent model also has smaller H$\alpha$,  and this  difference can be explained in terms of the H$\alpha$ source function. Most of the H$\alpha$ flux comes from the first 10 stellar radii XXX

Another potentially useful result is illustrated in Figure~\ref{fig:nir}, where we show the Bracket IR spectrum for each model.  
Fitting the continuum spectrum with a power-law gives us the following spectral indices: $-3.003\pm0.001$ for the consistent model; $-3.167\pm0.001$ for the mixed model; $-3.144\pm0.001$ for the isothermal model.
The NIR spectral index probes the density structure of the inner disk. 
As expected, the spectral indices of the mixed and isothermal models are nearly identical, since they share the same density structure. The NIR spectrum for the consistent model, however, has a lower spectral index because of the changes of the radial density structure of the inner disk. This suggests the interesting prospect of probing the temperature structure of the inner disk by way of detailed observations of the NIR continuum spectrum. 

\section{Conclusions}
We have investigated how the non-isothermal temperatures alter the density structure of Keplerian viscous disks with steady-state outflow.  For low mass loss rates (low densities) there is relatively little affect on the disk observables because the only changes occur in the innermost region of the disk where it is optically thick.  However for high mass loss rates, the disk is optically thick out to a much larger radius, so the effects are much more pronounced.

The primary effect is that the temperature decrease causes the disk to collapse, becoming much thinner in the inner regions.  This collapse redistributes the disk material toward the equator, increasing the mid-plane density by about a factor of 3 (relative to the isothermal density model). 
As a result, the disk opening angle is relatively constant (about $2$--$3\deg$) out to the location of the temperature minimum.  Beyond this location the temperature rise causes the disk to flare rapidly, and the disk opening angle approaches that of the isothermal model, which has opening angles of about $10^\circ$ at 20 stellar radii.

The combination of the radial temperature structure, disk scale height (opening angle), and viscous transport produces a complex radial dependence for the disk density.  The equivalent power law radial density exponent varies between $n=-2$ in the inner disk to $n=-5$ near the temperature minimum, eventually rising back to the isothermal value $n=-3.5$ in the outer disk.
  
The net result of these changes in the disk density (which alters the hydrogen level populations) is that the H$\alpha$ emission strength varies by about 
50\%, while the instrinsic polarization level changes by a factor of 2.  The effect on the spectral energy distribuion (SED) is somewhat smaller.  The IR continuum level changes by about 10\% with slight changes in the spectral index (slope).

We conclude that detailed modeling of the observations of Be star disks, such as the interpretation of interferometric, spectroscopic, and polarimetric observations, all require the use of the non-isothermal density structure.  
 
\acknowledgments{
This work was supported by FAPESP grant 04/07707-3 (ACC) and NSF   
grant AST-0307686 to the University of Toledo (JEB). 
}

\clearpage

%%%%%%%%%%%%%%%%%%%%%%%%%%
\begin{figure}[bp]
\plotone{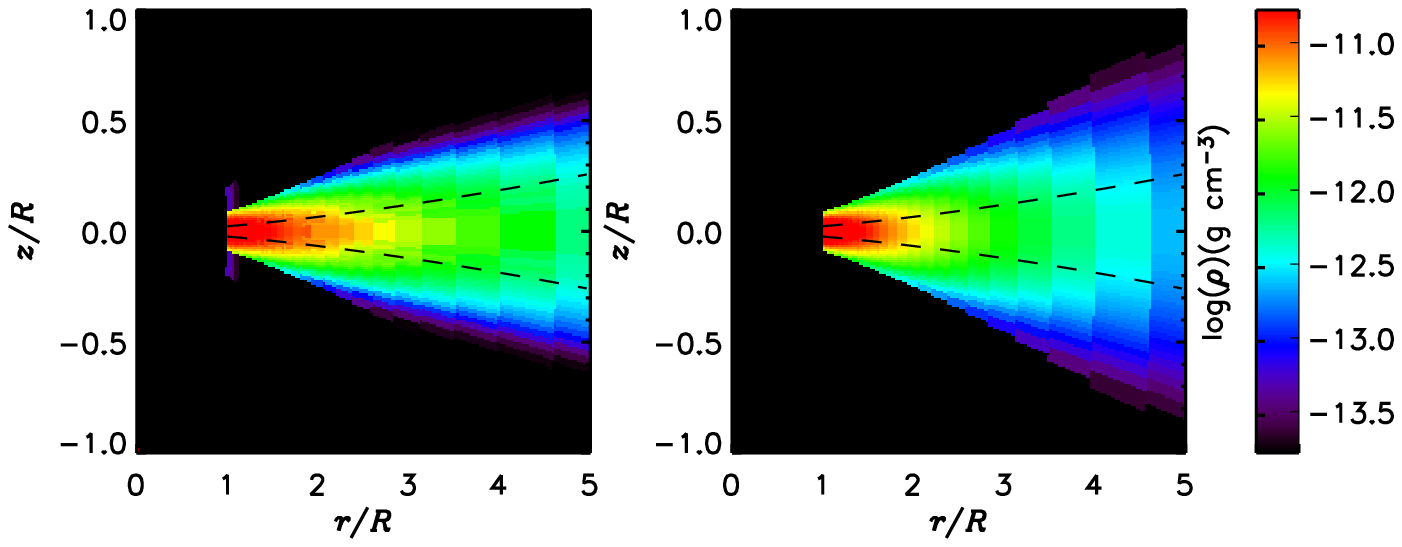}
\figcaption[]{Left: Solution for the disk structure of the high mass loss rate model of Table~\ref{tab:models}.
Right: density structure of a corresponding isothermal disk.
\label{fig:density}}
\end{figure}

%%%%%%%%%%%%%%%%%%%%%%%%%%
\begin{figure}[bp]
\centerline{\plotone{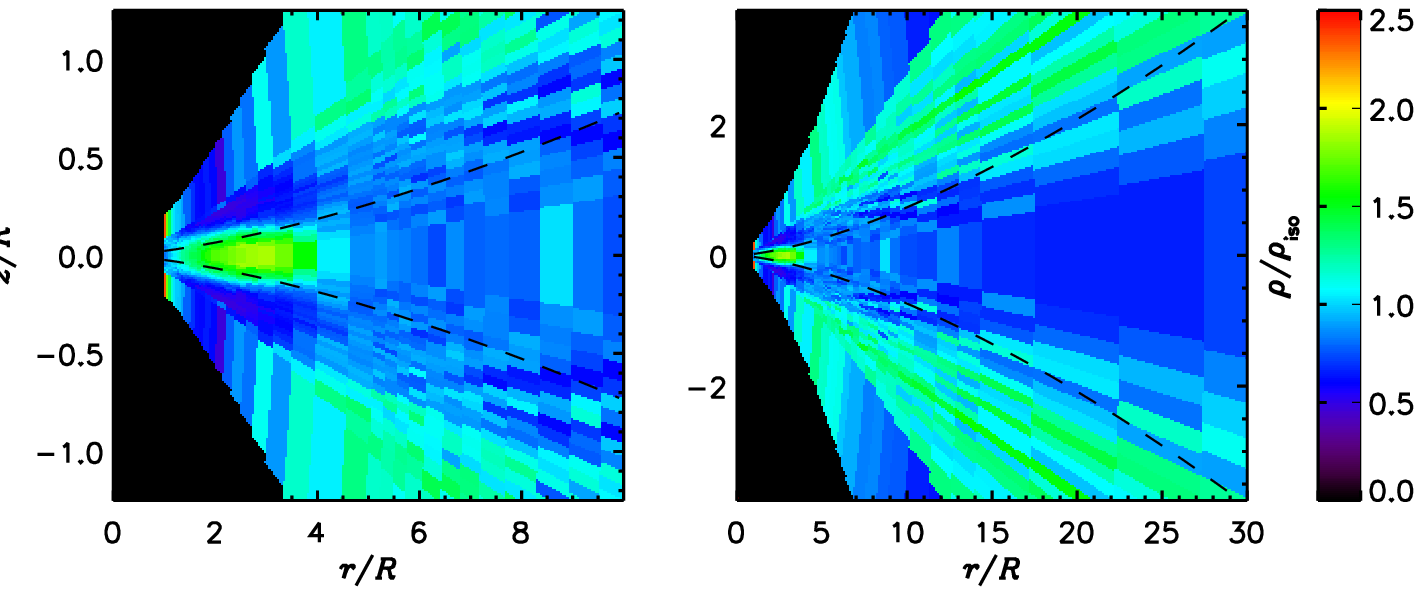}}
\centerline{\plotone{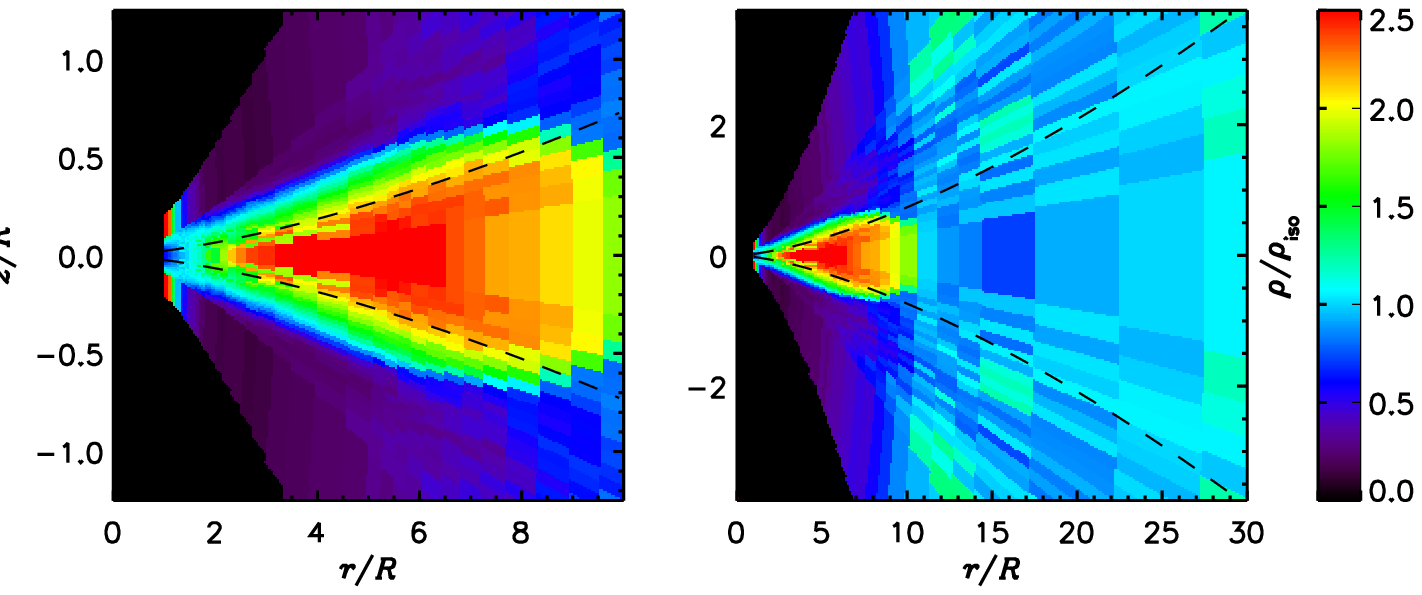}}
\figcaption[]{Comparison of the hydrostatic solution with the isothermal solution. Shown is the ratio  between the two solutions, $\rho/\rho_\mathrm{iso}$. The top panels correspond to the low mass loss rate model and the bottom panels to the high mass loss rate models.
The dashed lines correspond to the curves 
$|z| = H_{\rm isot}(\varpi)$.
\label{fig:deltarho}}
\end{figure}

%%%%%%%%%%%%%%%%%%%%%%%%%%
\begin{figure}[bp]
\centerline{\plottwo{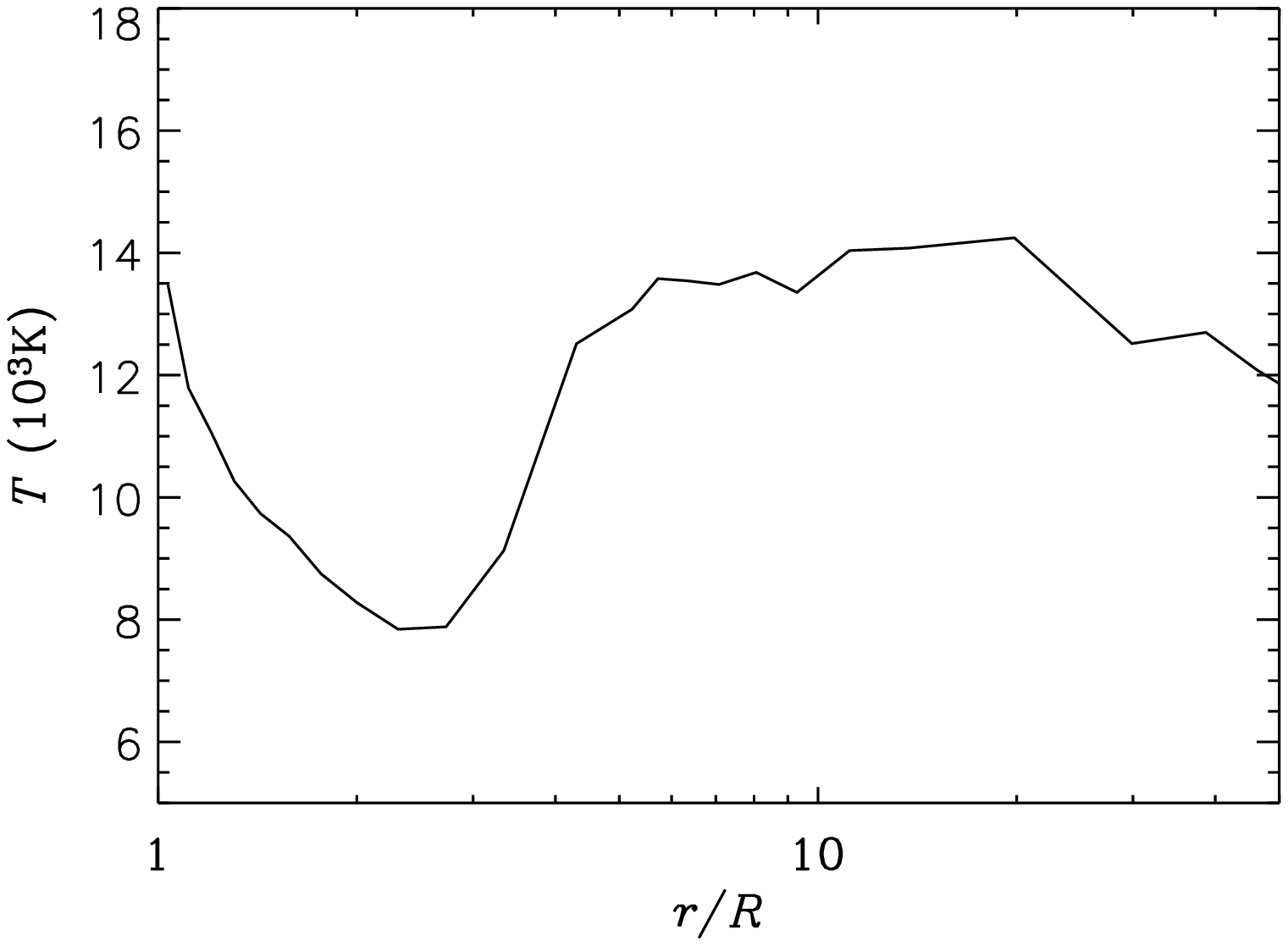}{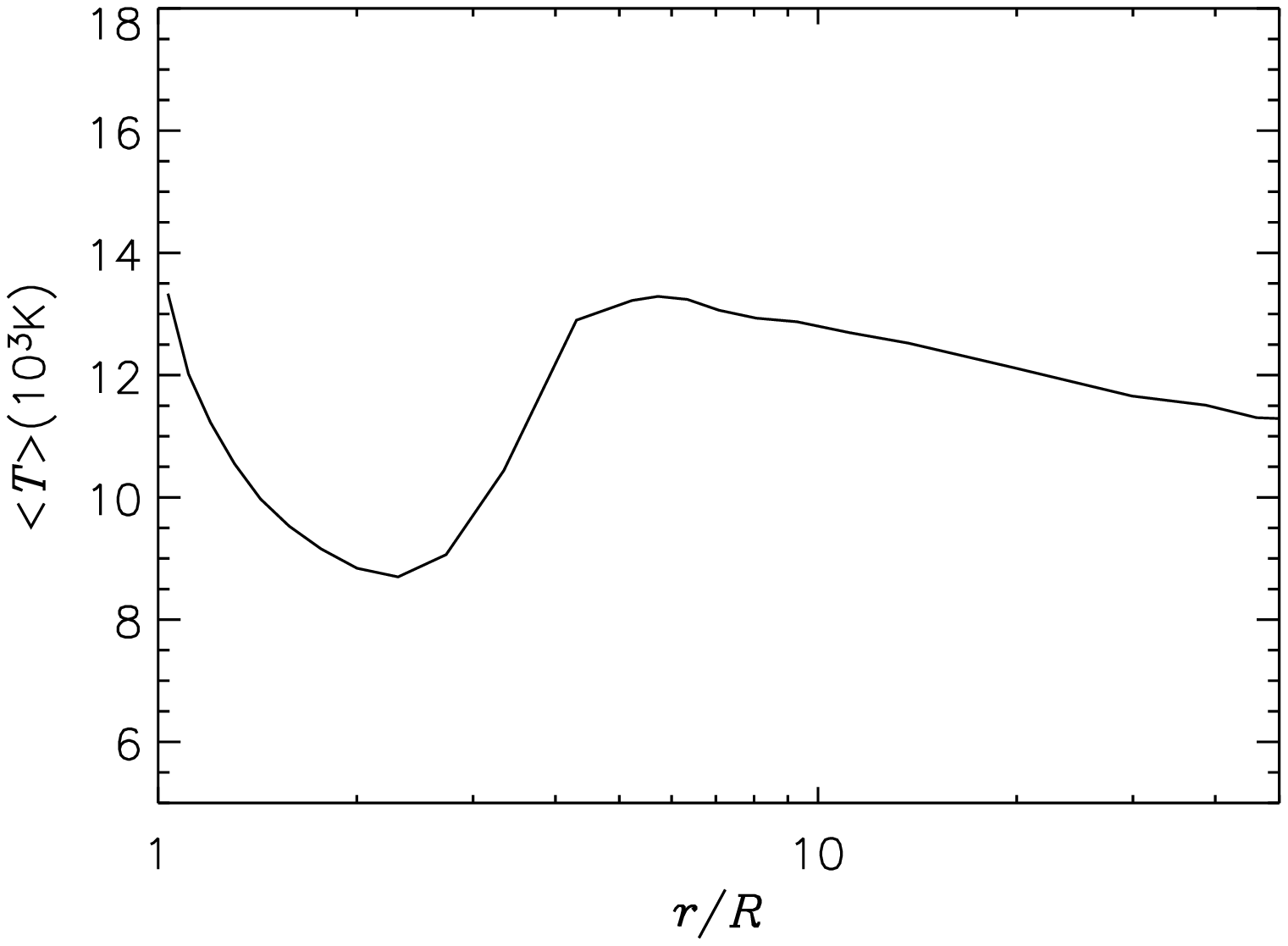}}
\centerline{\plottwo{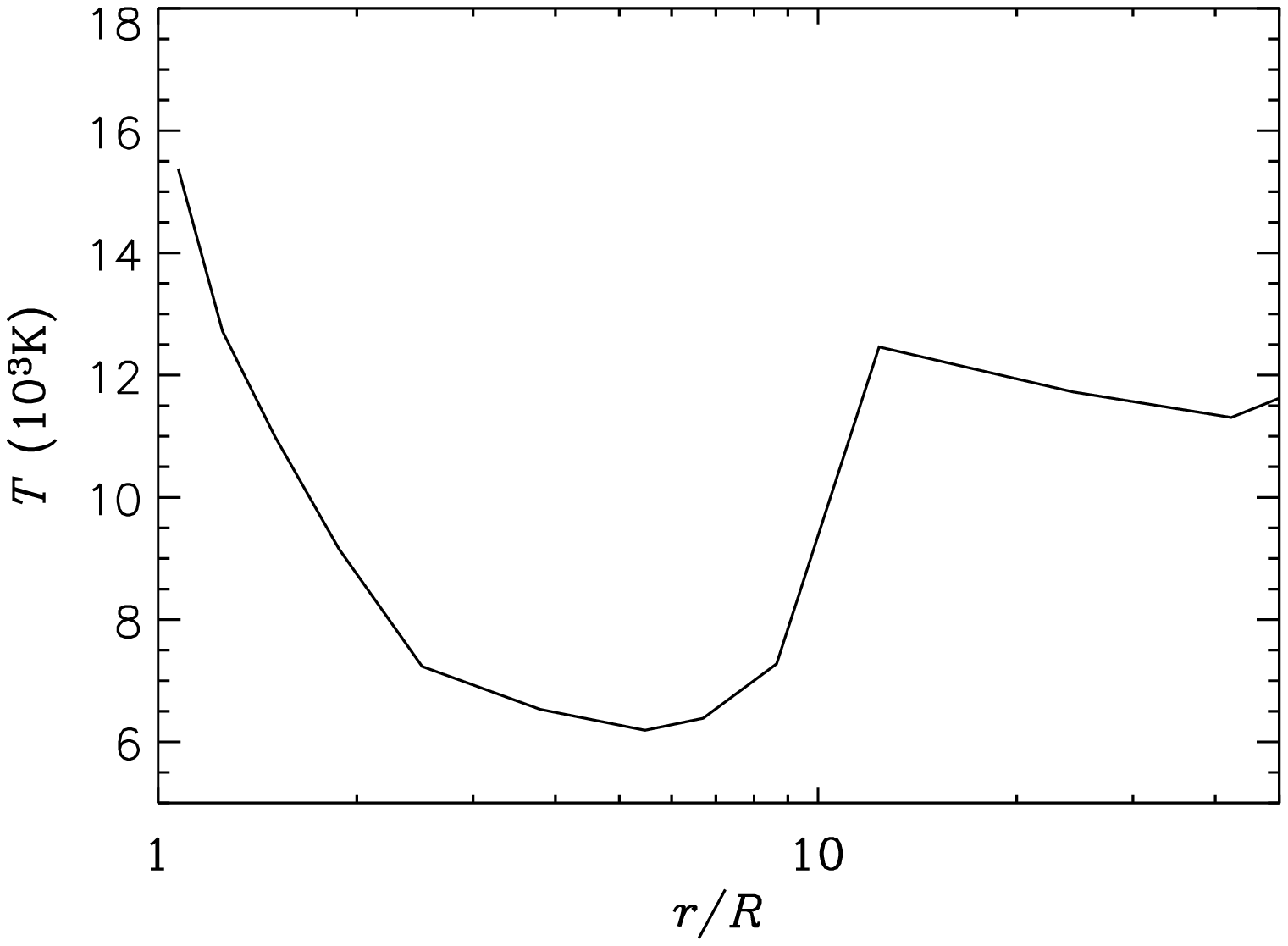}{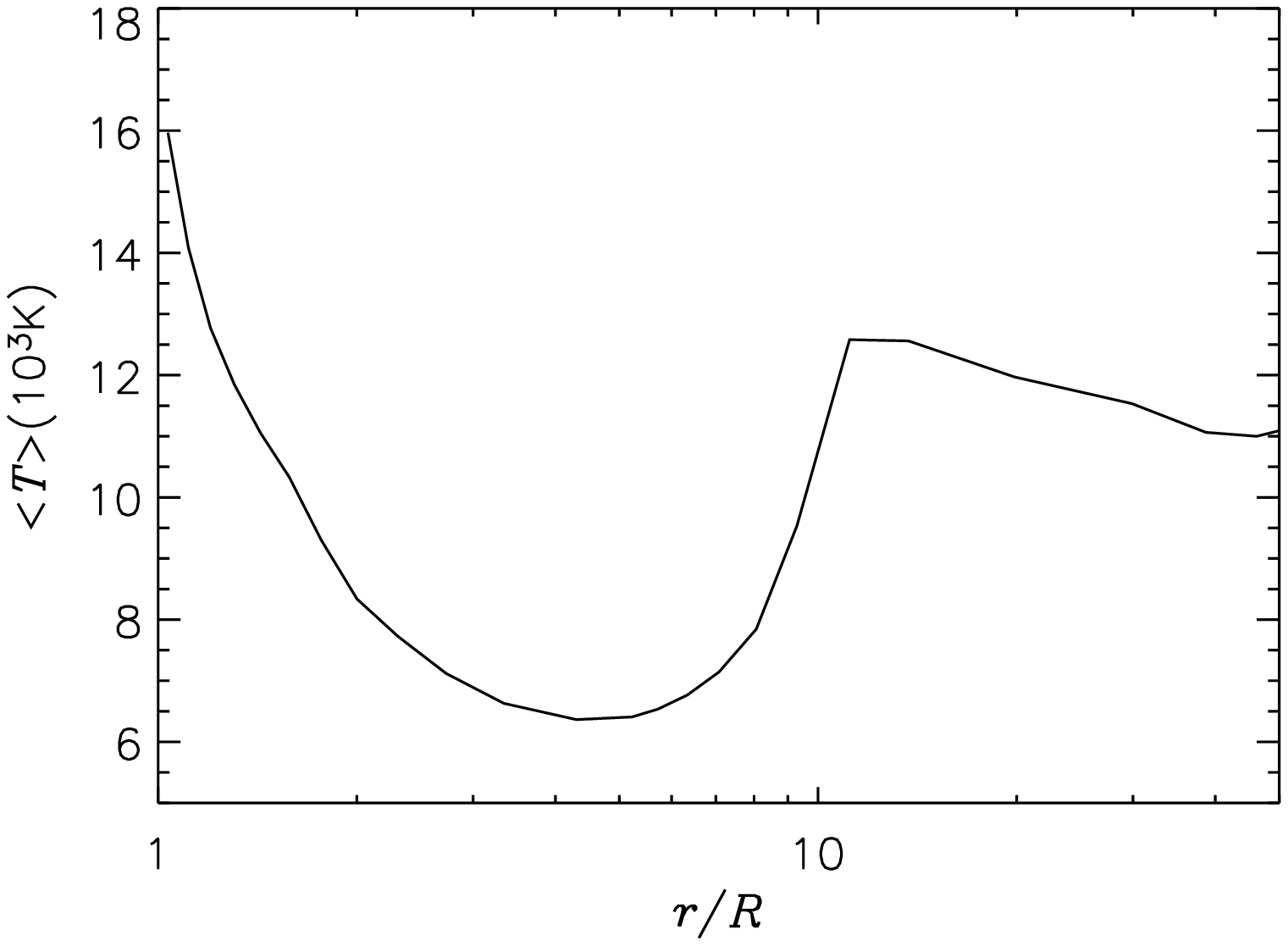}}
\caption{Radial temperature structure. 
Left panels: Temperature along the mid-plane. 
Right panels: Weighted-average of T in the vertical direction, eq.~(\ref{eq:avT}). 
The top panels correspond to the low mass loss rate model and the bottom panels to the high mass loss rate model.
}
 \label{fig:temperature}
\end{figure}

%%%%%%%%%%%%%%%%%%%%%%%%%%
\begin{figure}[6cm]
\epsscale{.9}
\centerline{\plottwo{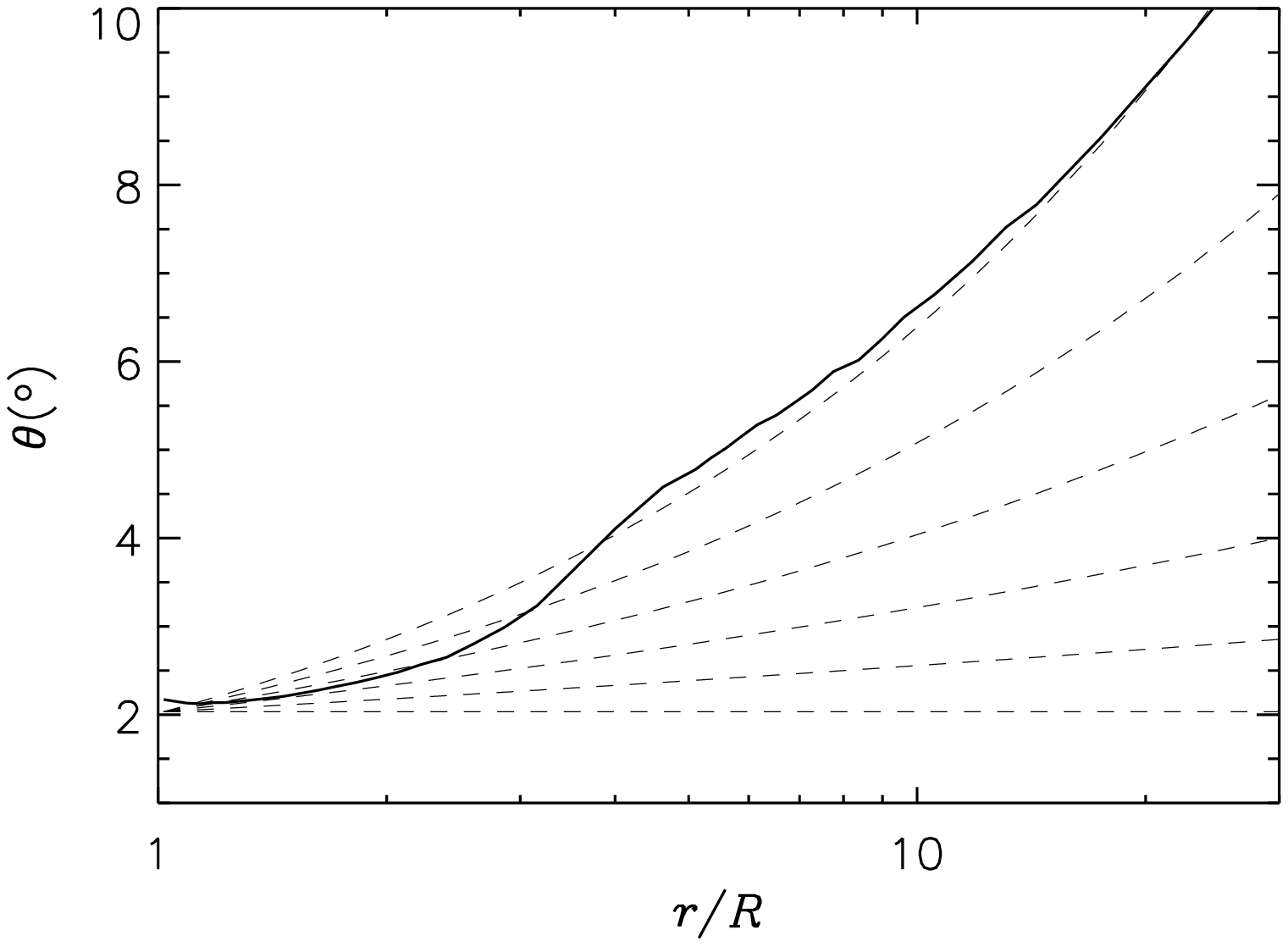}{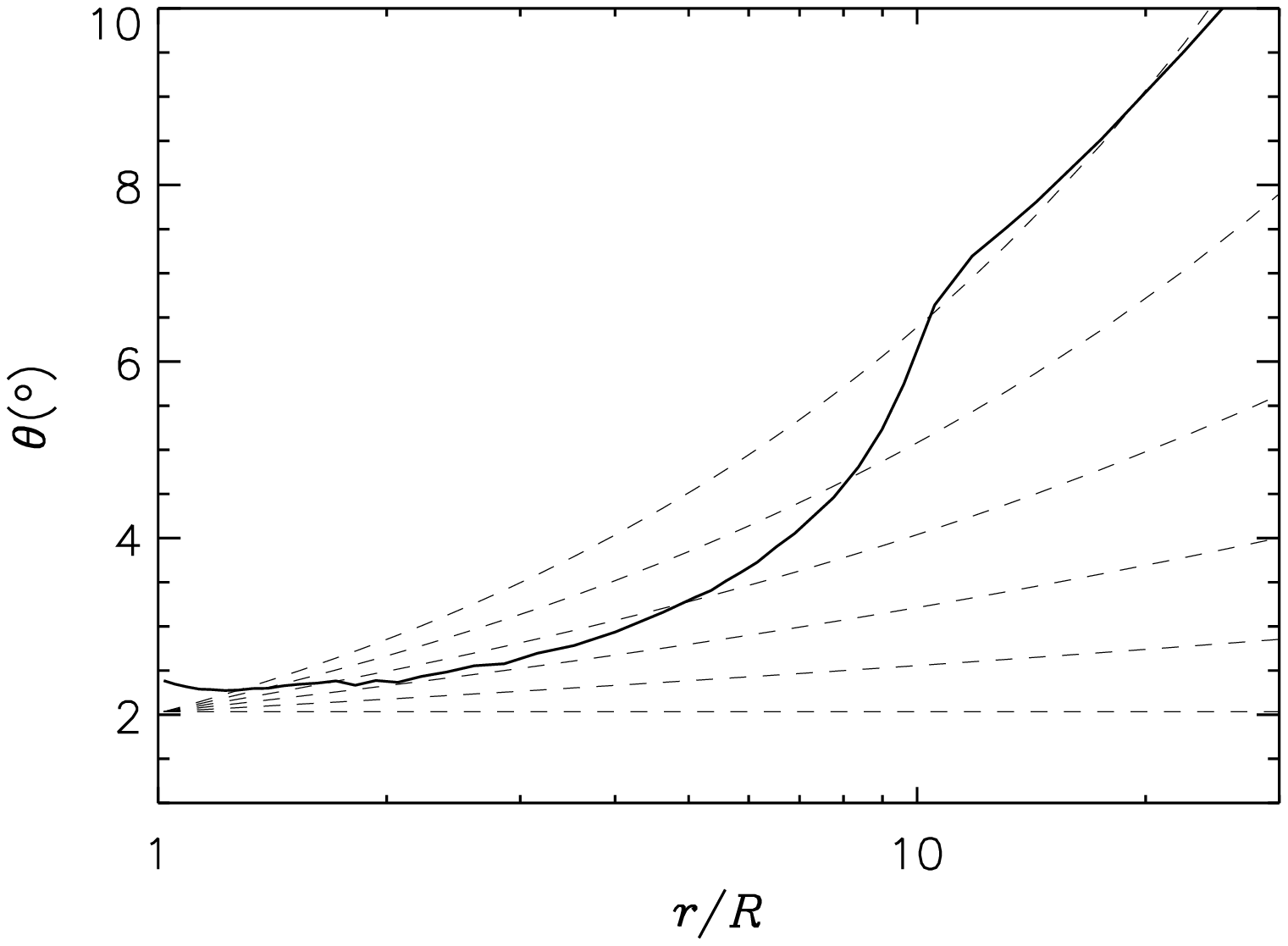}}
\centerline{\plottwo{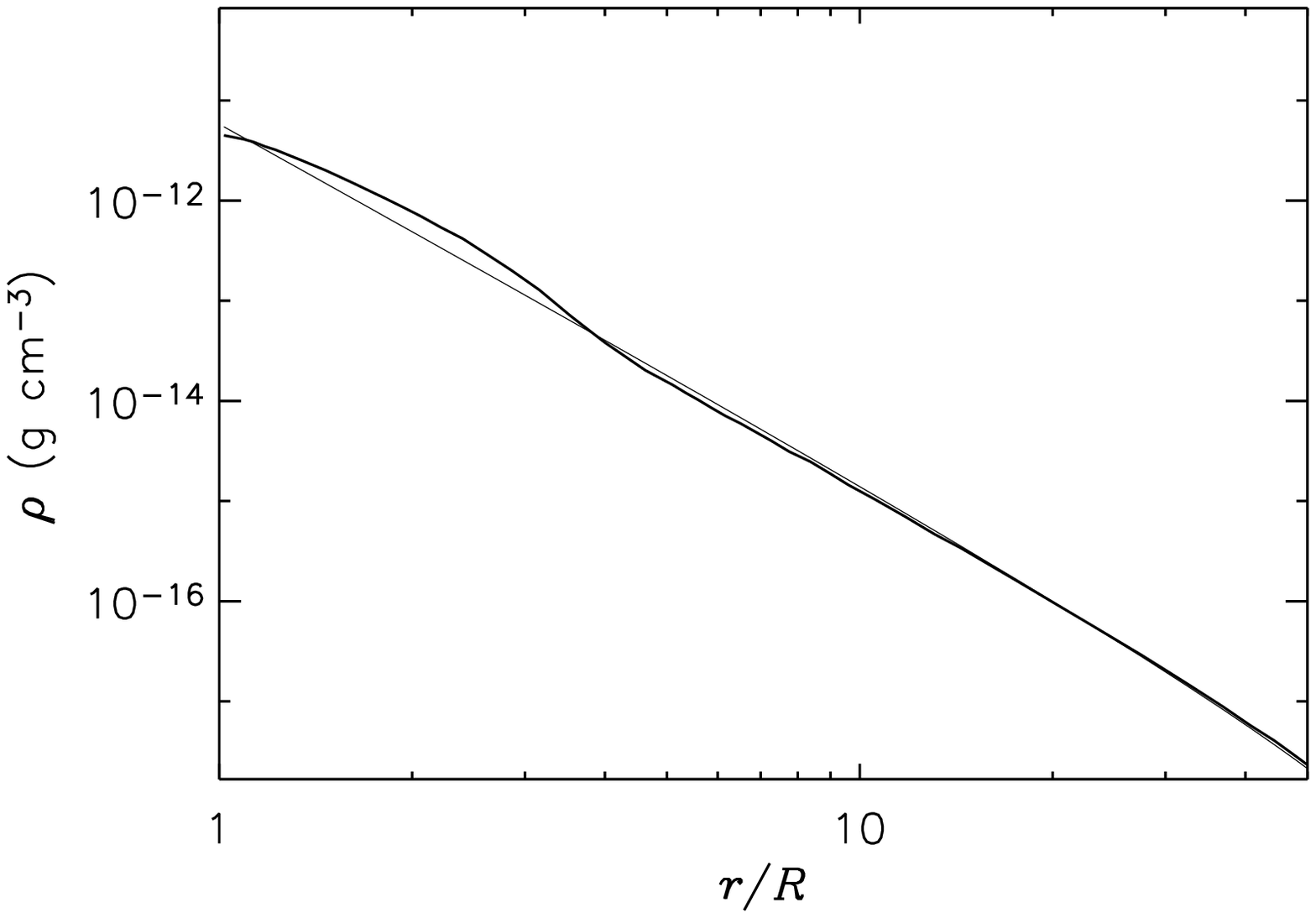}{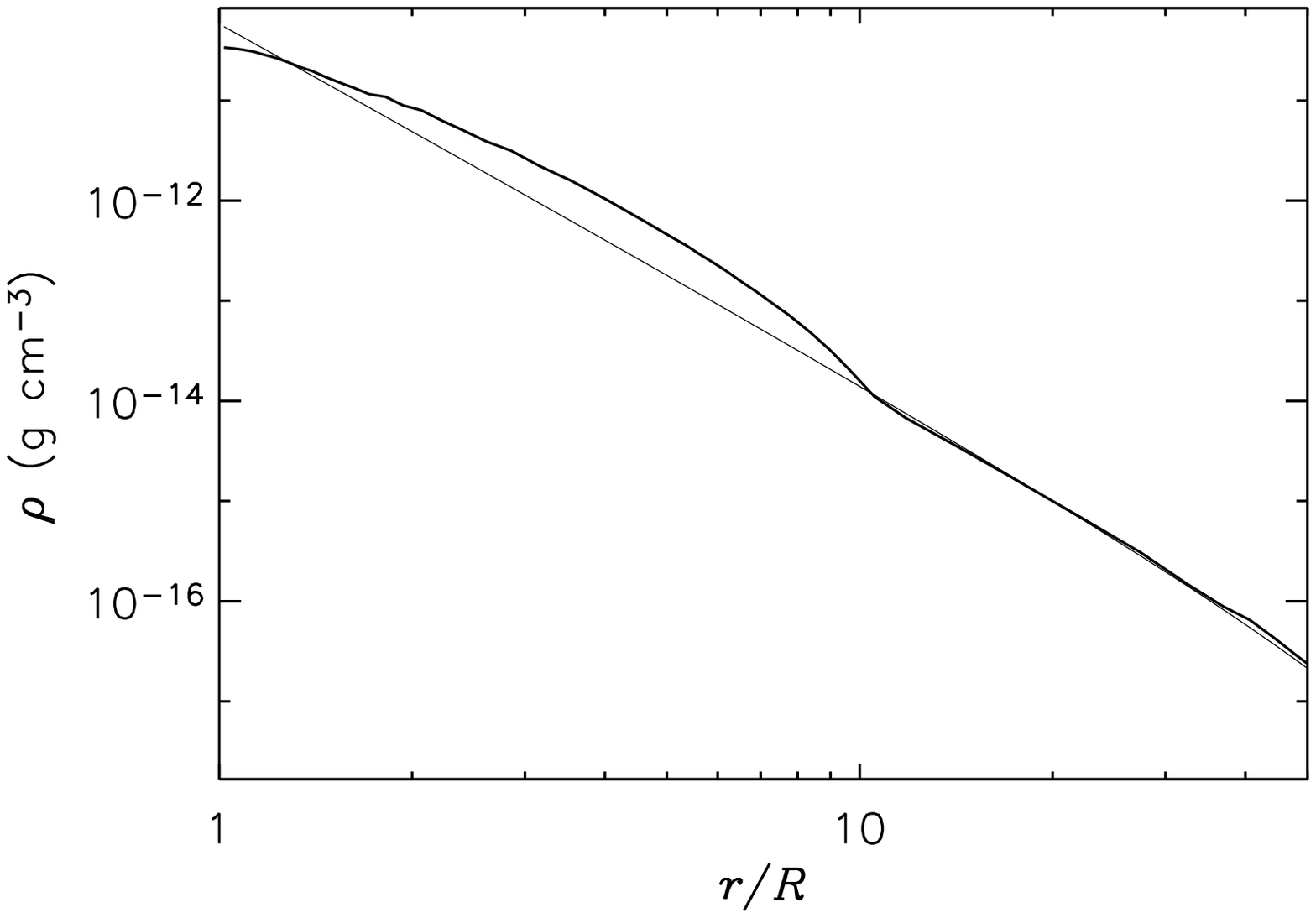}}
\centerline{\plottwo{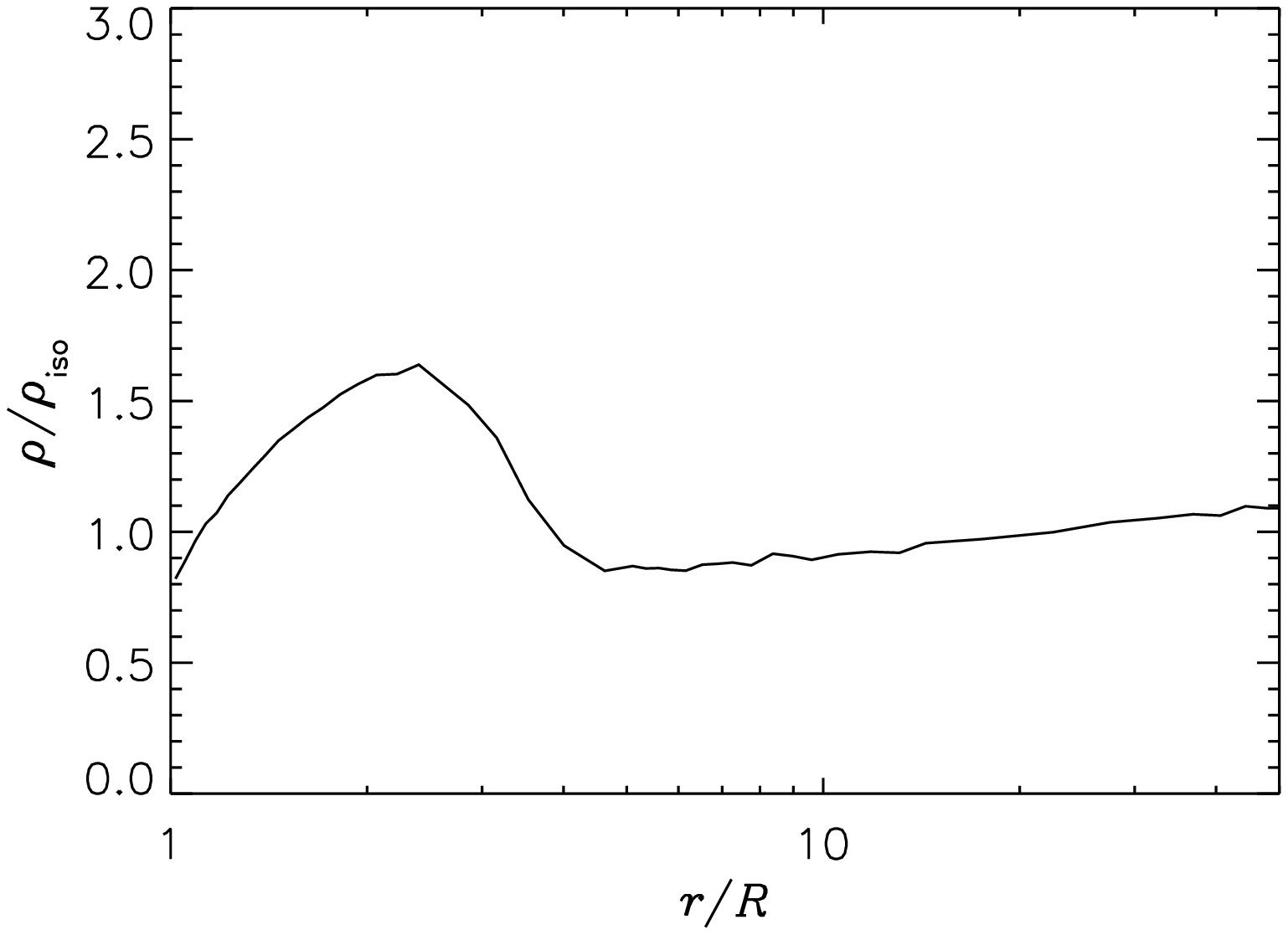}{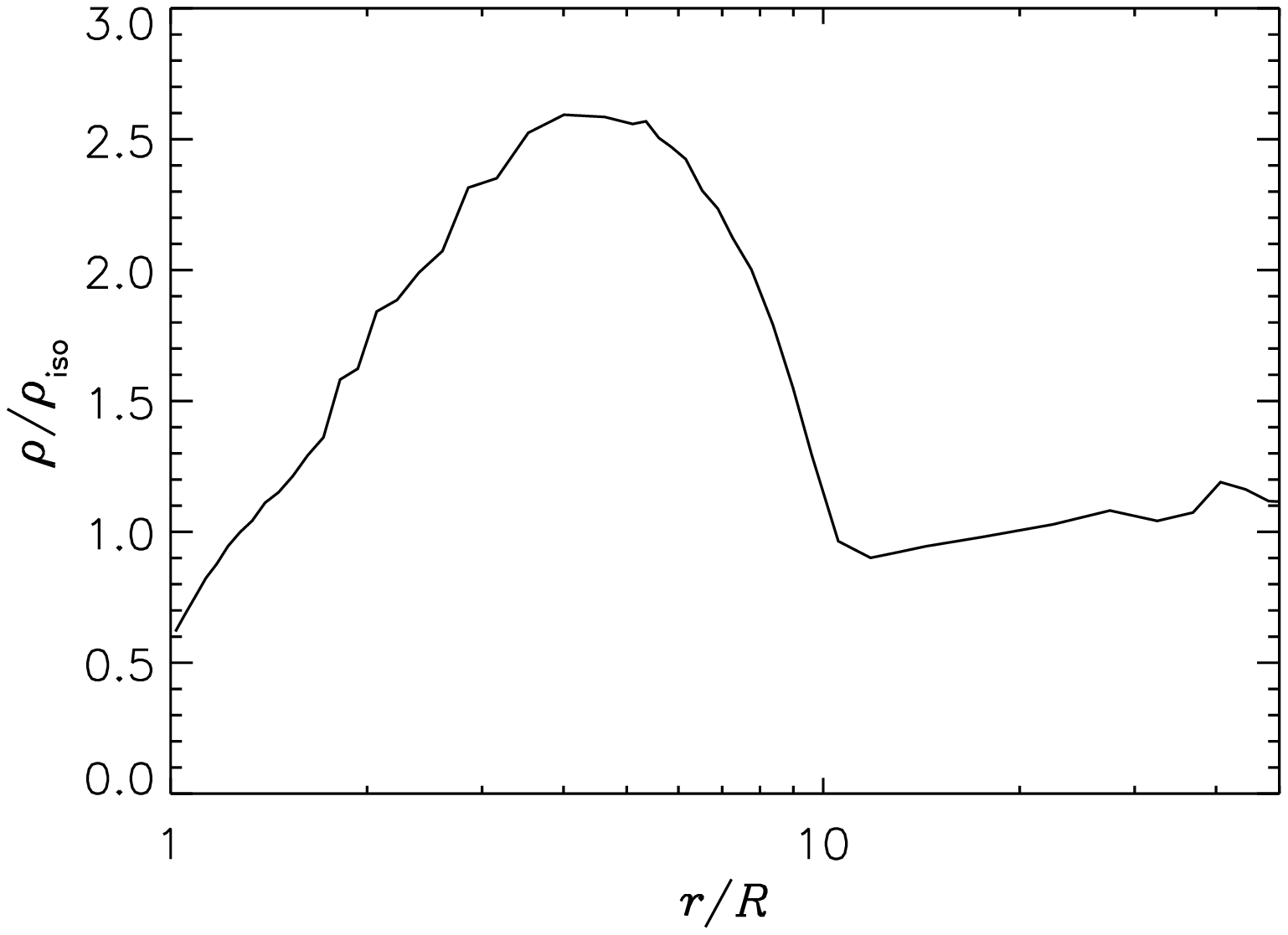}}
\centerline{\plottwo{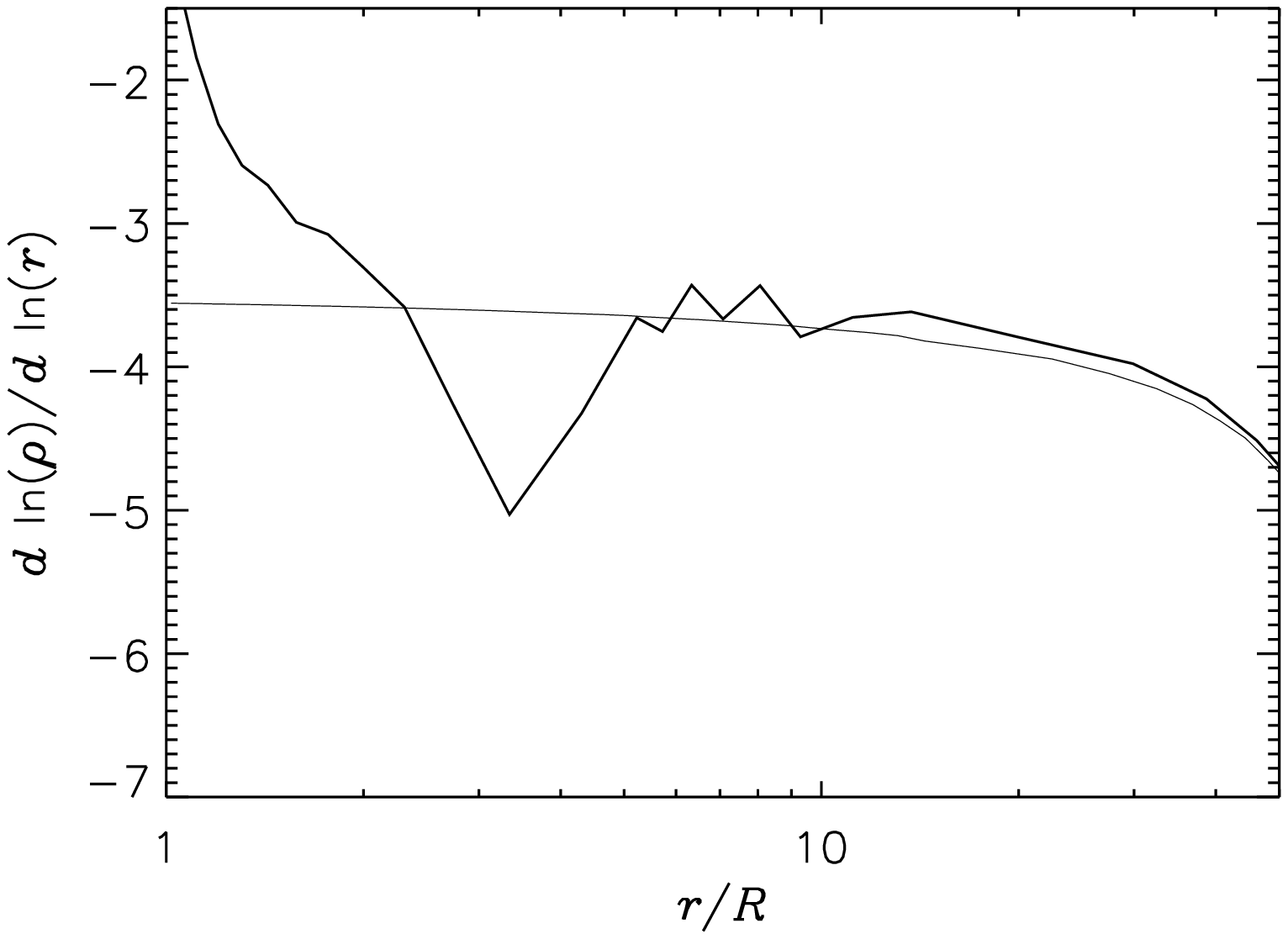}{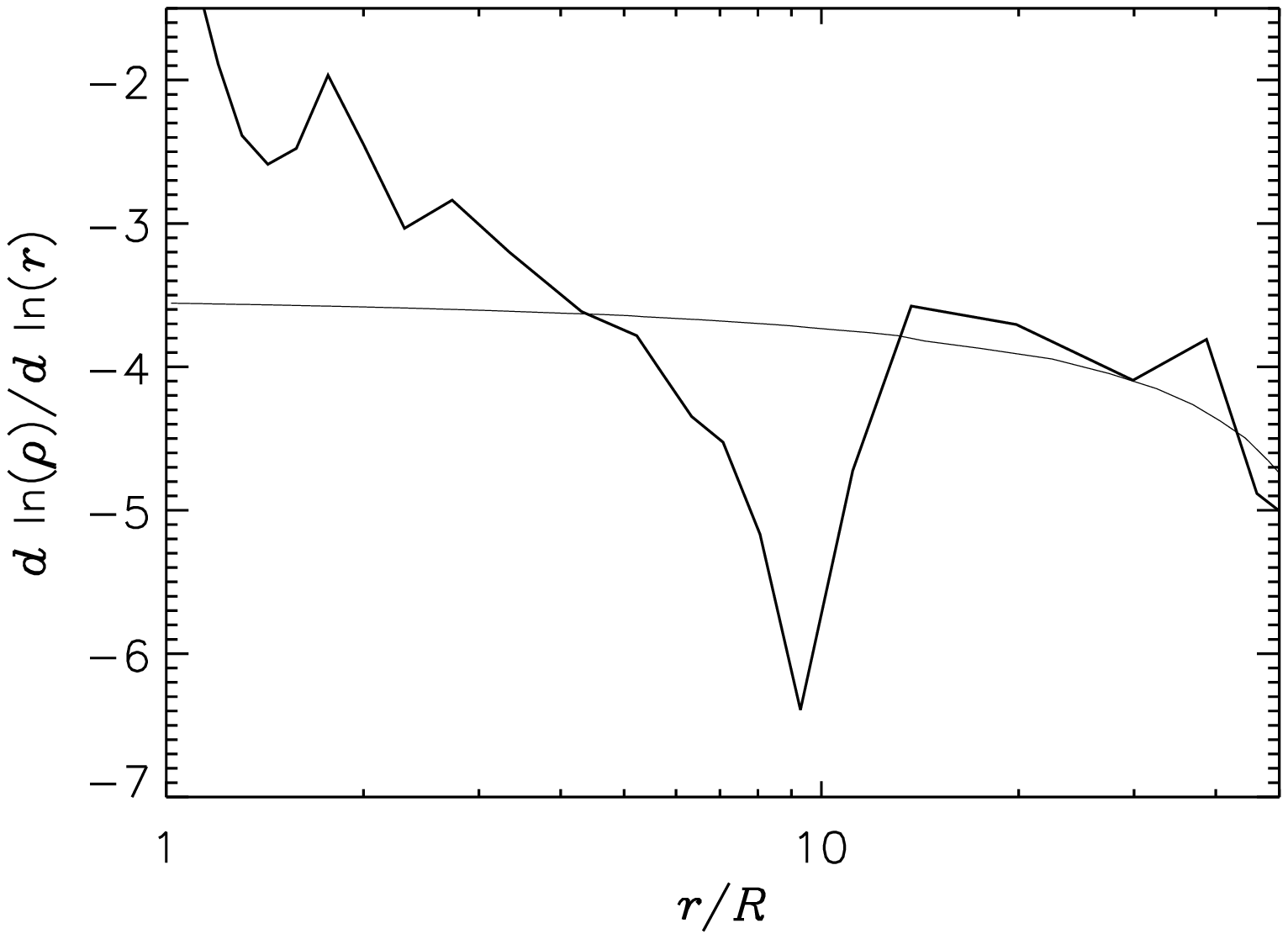}}
\epsscale{1}
\caption{Disk density structure of low (left) and high mass loss rate models (right).
Top: Opening angle. The thick line corresponds to the self-consistent solution for the disk opening angle [eq.~(\ref{eq:opangle})]. The dashed lines correspond to opening angles for scale-heights parameterized as $H(\varpi) \propto \varpi^\beta$; from bottom to top, $\beta=1.0$ to 1.5, in steps of 0.1. 
Second from top: Density profile. We compare the consistent density calculated by HDUST (thick line) with the isothermal density (thin line).
Second from bottom: Ratio between the consistent density and the isothermal density.
Bottom: Local index of the density profile. 
}
 \label{fig:structure}
\end{figure}

%%%%%%%%%%%%%%%%%%%%%%%%%%
\begin{figure}[bp]
\plotone{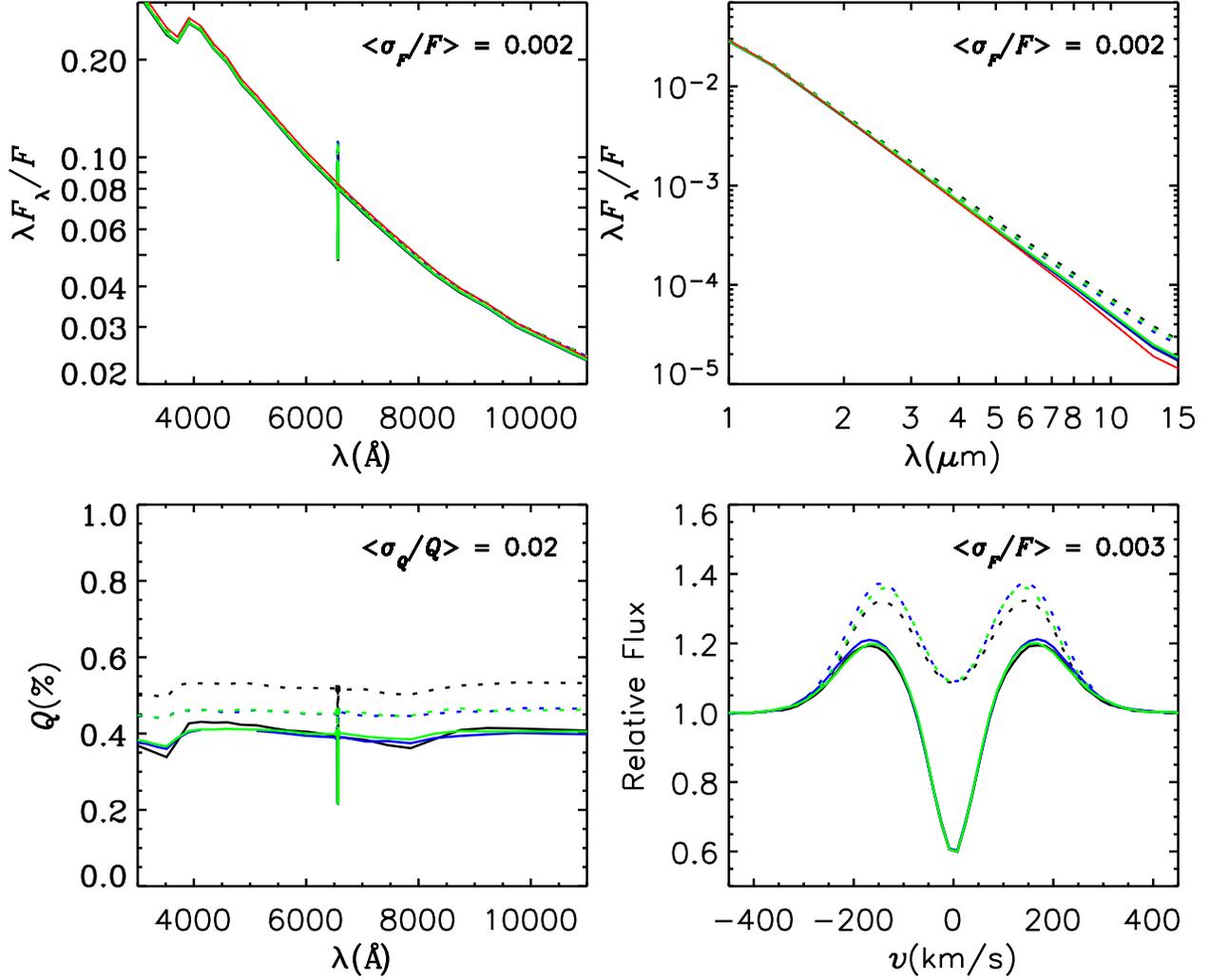}
\figcaption[]{Emergent spectrum for the low mass loss rate model of Table~\ref{tab:models} (black lines). 
The solid lines correspond to a viewing angle of 90\degr (disk seen edge-on) and the dotted lines to 70\degr.
Top panels: SED. Bottom left: polarization. Bottom right: H$\alpha$ profile (a flat stellar profile was assumed). For comparison, we show in blue the results for the mixed model and in green the results for the isothermal model. For the latter models we adopted an electron temperature of 60\% of $T_\mathrm{eff}$.
The red curves correspond to the stellar SED, without the disk.
We indicate in each panel the average error of the Monte Carlo simulation, that ranges from $0.2$ --- $2\%$.
\label{fig:mosaic1}}
\end{figure}

%%%%%%%%%%%%%%%%%%%%%%%%%%
\begin{figure}[bp]
\plotone{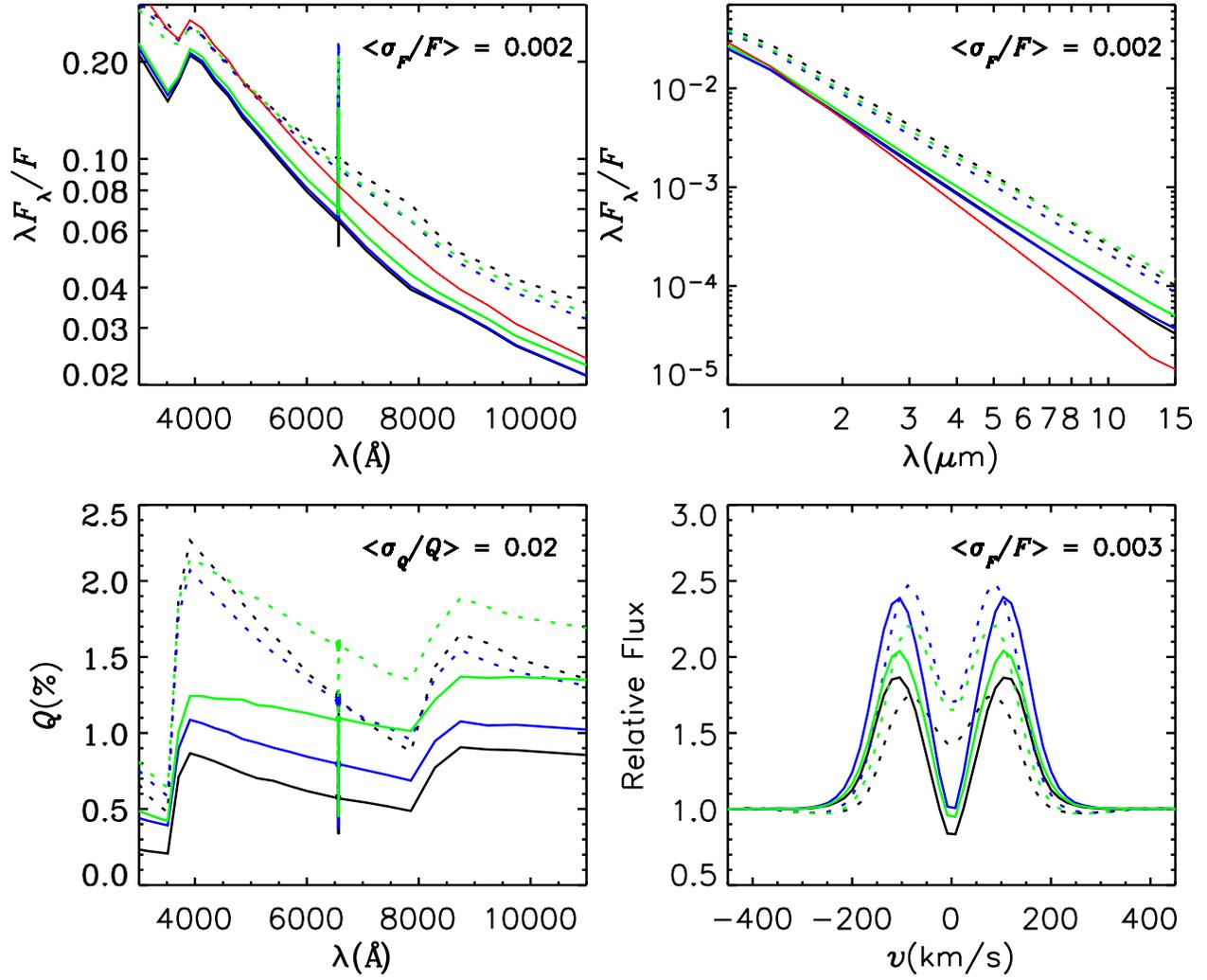}
\figcaption[]{Same as Fig.~\ref{fig:mosaic1} for the equal mass  model of Table~\ref{tab:models}.
\label{fig:mosaic2}}
\end{figure}

%%%%%%%%%%%%%%%%%%%%%%%%%%
\begin{figure}[bp]
\plotone{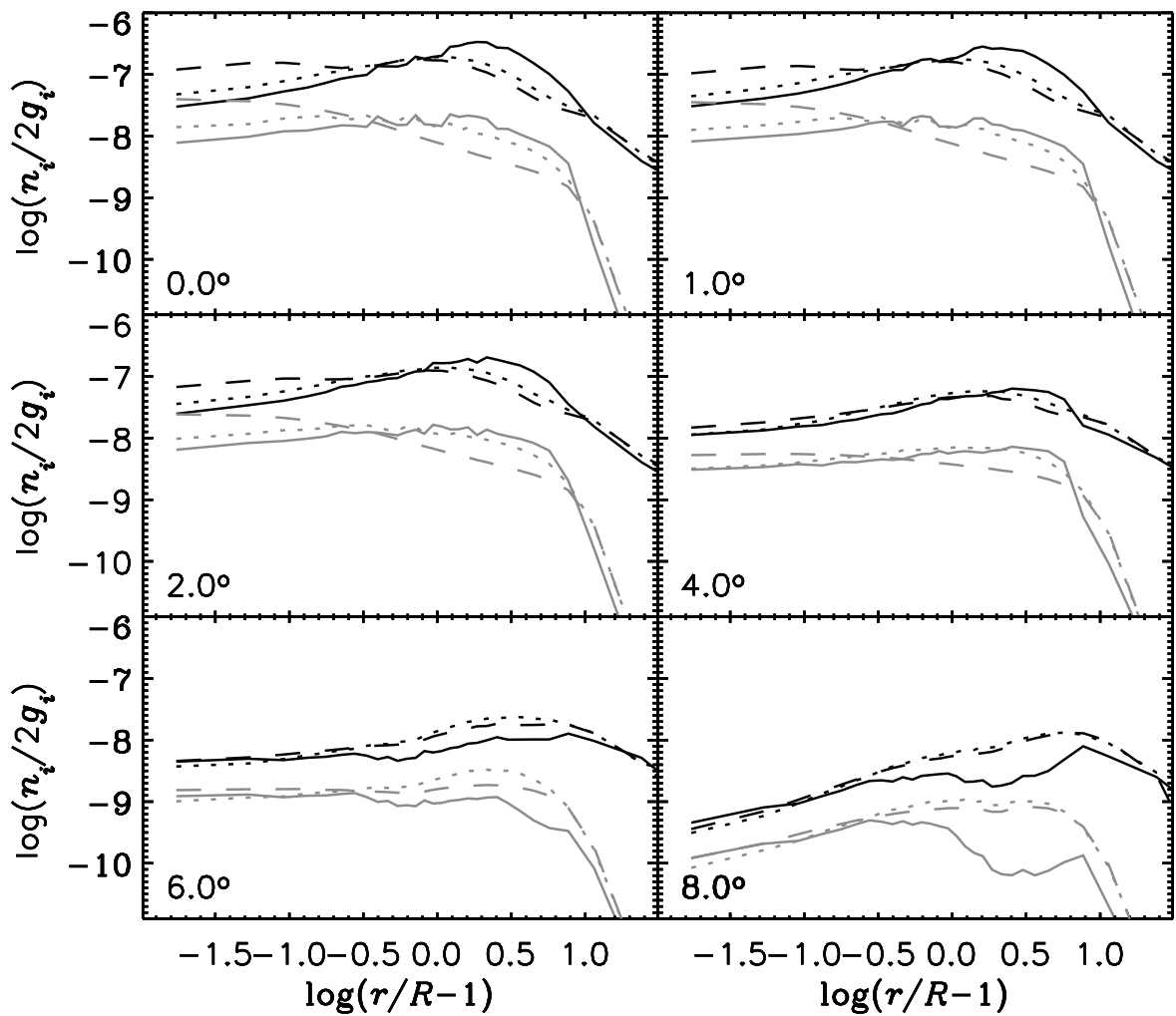}
\figcaption[]{
{Hydrogen level populations for the high mass loss rate case. Shown are the level populations for $n=2$ (black), and $n=3$ (gray) as a function of radius. 
Solid lines: consistent model. Dotted lines: mixed model. Dashed lines: isothermal model.
Each panel shows the populations at a given latitude (angle above mid-plane), as indicated.
}
\label{fig:lev}}
\end{figure}

%%%%%%%%%%%%%%%%%%%%%%%%%%
\begin{figure}[bp]
\plotone{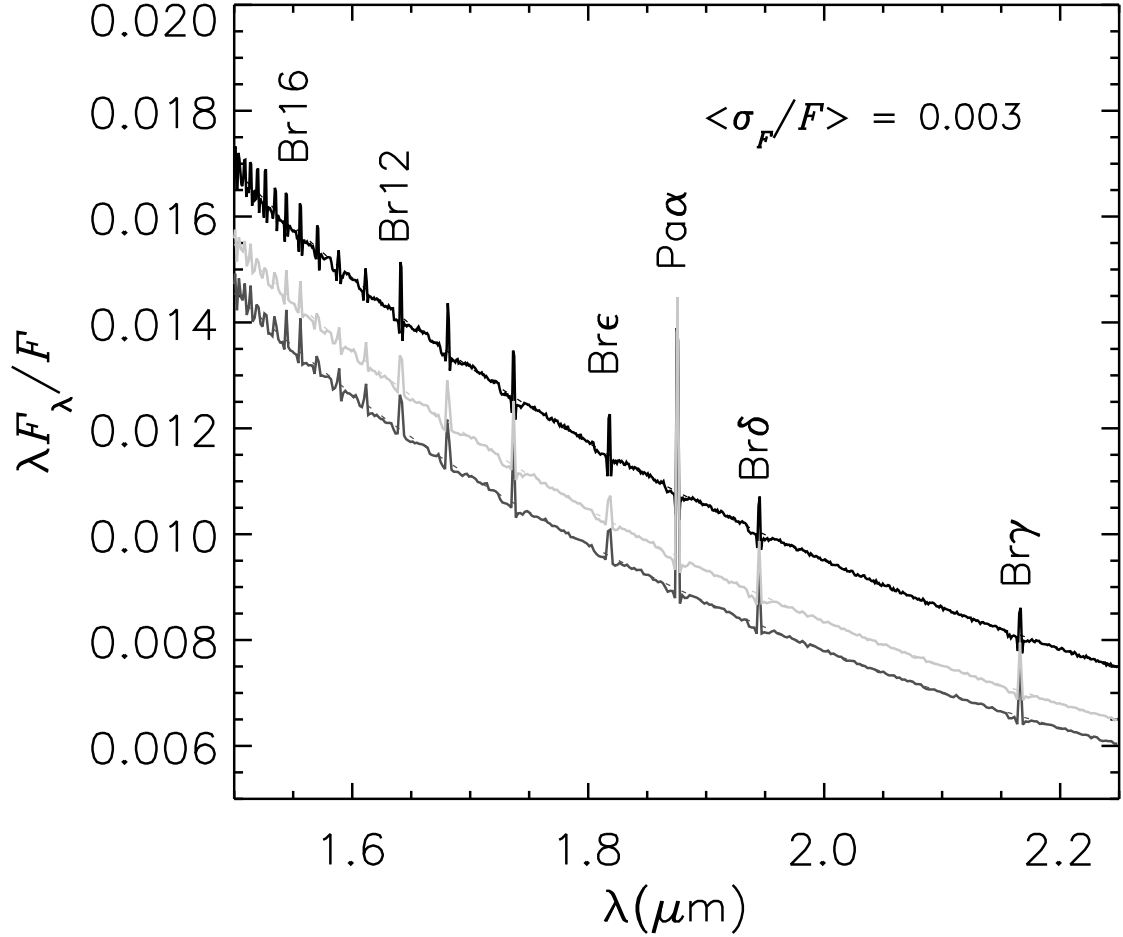}
\figcaption[]{Bracket continuum for the high mass loss rate models of Table~\ref{tab:models} (black line).
For comparison, we show the mixed (dark gray) and isothermal (light gray) models. A power-law fit of the continuum flux is also shown as dashed lines.
The average error of the Monte Carlo simulation is $0.3\%$.
\label{fig:nir}}
\end{figure}

\clearpage

%%%%%%%%%%%%%%%%%%%%%%%%%%%%%%%%%%%%%%%%%%%%%%%%%%%
\clearpage
\begin{deluxetable}{ccccc}
\tablecaption{Fixed Model Parameters \label{tab:fixed}}
\tablewidth{0pt}
\tablehead{
\colhead{$R$} &
\colhead{$T_\mathrm{eff}$} &
\colhead{$V_\mathrm{crit}$} &
\colhead{$\alpha$} &
\colhead{$R_{\rm d}$}
\\
\colhead{$(R_{\sun})$} & 
\colhead{(K)} & 
\colhead{(km\,s$^{-1}$)} & 
\colhead{} & 
\colhead{$(R)$}
}
\startdata
7 & $20\,000$ & 400  & 0.1 & 100 \\
\enddata
\end{deluxetable}

%4,32E-12	5,48E-12
%2,99E-11	5,48E-11
%2,49E-11	0,00E+00
%%%%%%%%%%%%%%%%%%%%%%%%%%%%%%%%%%%%%%%%%%%%%%%%%%%
\begin{deluxetable}{lccc}
\tablecaption{Model Densities \label{tab:models}}
\tablewidth{0pt}
\tablehead{
\colhead{} &
\colhead{$\dot{M}$} &
\colhead{$\rho_0(\varpi=R)$} &
\colhead{$\rho_0(\varpi=R)$}\tablenotemark{a}
\\
\colhead{} & 
\colhead{$(M_{\sun}\;\rm yr^{-1})$} &
\colhead{$(\rm g\;cm^{-3})$} &
\colhead{$(\rm g\;cm^{-3})$}
}
\startdata
Low mass loss rate & $5.0\times 10^{-12}$ & $4.3\times10^{-12}$ & $5.5\times10^{-12}$ \\
Equal Mass & $4.9\times 10^{-12}$ & $4.2\times10^{-12}$ &$5.4\times10^{-12}$ \\
High mass loss rate & $5.0\times 10^{-11}$ & $3.0\times10^{-11}$ & $5.5\times10^{-11}$\\
Equal Mass & $3.9\times 10^{-11}$ & $2.5\times10^{-11}$ &$4.3\times10^{-11}$ \\
\enddata
\tablenotetext{a}{Value for an isothermal model}
\end{deluxetable}

\end{document}